%% file: EIG_-_I._The_Sample.tex
\documentclass[useAMS,usenatbib]{mn2e}
\usepackage{graphicx,rotating,subfigure,amsmath,wasysym,amssymb}
\usepackage[retainorgcmds]{IEEEtrantools}
\usepackage{ctable}
\usepackage{multirow}

\input{mycommands.tex}

\input{journals.tex}
\title[EIG - I. Sample and Simulation Analysis]
      {Extremely Isolated Galaxies \\ I. Sample and Simulation Analysis}
\author[O. Spector and N. Brosch]
{O. Spector$^{}$\thanks{E-mail: odedspec@wise.tau.ac.il} and N. Brosch \\
$^{}$Wise Observatory and the Raymond and Beverly Sackler School of Physics and Astronomy, \\
       Tel Aviv University, Tel Aviv 69978, Israel\\
}

\begin{document}

\date{Accepted 2015 November 10. Received 2015 November 9; in original form 2015 July 5}

\pagerange{\pageref{firstpage}--\pageref{lastpage}} \pubyear{2016}

\maketitle

\label{firstpage}

\input{Chapters/Abstract}


\input{Chapters/1_Introduction}
\input{Chapters/2_TheSample}

\input{Chapters/3_Environment}

\input{Chapters/7_Conclusions}

\input{Chapters/acknowledgments}

\addcontentsline{toc}{chapter}{Bibliography}
\bibliographystyle{mn2e}
\bibliography{references}

\appendix
\input{Chapters/appB_IndividualGalaxies}

\label{lastpage}

\end{document}

%% file: mycommands.tex
\newcommand{\rem}[1]{}  

\newcommand{\markChange}[1]{{#1}} 

\newcommand  \dist[1]   {\ifmmode d_{#1}\else $d_{#1}$\fi} 

\def \log       {\ifmmode \mbox{log}\else log\fi}
\def \dex       {\ifmmode \mbox{dex}\else dex\fi}

\def  \sun	{\ifmmode \odot \else \mbox{$\odot$} \fi}
\def  \earth	{\ifmmode \oplus \else \mbox{$\oplus$} \fi}

\def  \Hubble   {\ifmmode H_{0}\else $H_{0}$\fi} 
\def  \h        {\ifmmode h\else $h$\fi} 

\def  \deg	{\ifmmode ^{\circ}\else        $^{\circ}$\fi}
\def  \arcmin	{\ifmmode ^\prime\else         $^\prime$\fi}
\def  \arcsec	{\ifmmode ^{\prime\prime}\else $^{\prime\prime}$\fi}
\def  \RA       {\ifmmode \alpha\else          $\alpha$\fi}  
\def  \Dec      {\ifmmode \delta\else          $\delta$\fi} 

\def  \sr       {\ifmmode sr\else              $sr$\fi}                     
\def  \sqDeg    {\ifmmode \mbox{deg}^{2}\else  $\mbox{deg}^{2}$\fi}         

\def  \yr       {\ifmmode \mbox{yr} \else $\mbox{yr}$ \fi} 
\def  \Myr      {\ifmmode \mbox{Myr}\else $\mbox{Myr}$\fi} 
\def  \Gyr      {\ifmmode \mbox{Gyr}\else $\mbox{Gyr}$\fi} 

\def  \GHz      {\ifmmode \mbox{GHz} \else GHz \fi} 

\def  \um      {\ifmmode \mu\!\mbox{m}\else $\mu\!\mbox{m}$\fi} 
\def  \nm      {\ifmmode \mbox{nm}\else nm\fi} 
\def  \Angst   {\ifmmode \mbox{\AA}\else  $\mbox{\AA}$\fi} 
\def  \pc      {\ifmmode \mbox{pc}  \else $\mbox{pc}$ \fi} 
\def  \kpc     {\ifmmode \mbox{kpc}\else $\mbox{kpc}$\fi} 
\def  \Mpc     {\ifmmode \mbox{Mpc}\else $\mbox{Mpc}$\fi} 

\def  \Mpch    {\ifmmode \h^{-1}\,\Mpc\else $\h^{-1}\,\Mpc$\fi} 
\def  \VolMpch {\ifmmode \h^{-3}\,\Mpc^{3}\else $\h^{-3}\,\Mpc^{3}$\fi} 
\def  \NumCntMpch {\ifmmode \h^{3}\,\Mpc^{-3}\else $\h^{3}\,\Mpc^{-3}$\fi} 

\def  \kms	{\ifmmode {\rm km\,s}^{-1}\else {km\,s$^{-1}$}\fi}

\def  \z       {\ifmmode \mbox{z}\else  z\fi}  
\def  \cz      {\ifmmode \mbox{cz}\else cz\fi} 

\def  \Mass	 {\ifmmode M                \else $M$                 \fi}
\def  \Mearth	 {\ifmmode \Mass_\earth     \else $\Mass_\earth$      \fi}
\def  \Msun      {\ifmmode \Mass_\sun\else        $\Mass_\sun$\fi}
\def  \Msunh     {\ifmmode \h^{-1}\,\Msun\else    $\h^{-1}\,\Msun$\fi}      
\def  \MHI       {\ifmmode \Mass_{HI}\else        $\Mass_{HI}$\fi}
\def  \Mstar     {\ifmmode \Mass_{*}\else         $\Mass_{*}$\fi}  
\def  \Mhalo     {\ifmmode \Mass_{halo}\else      $\Mass_{halo}$\fi} 

\def  \density         {\ifmmode \rho        \else $\rho$        \fi} 
\def  \densityMsunMpch {\ifmmode {\h^{2}\,\Msun\,\Mpc^{-3}}\else 
                                 $\h^{2}\,\Msun\,\Mpc^{-3}$ \fi} 

\def  \Lunimosity {\ifmmode L \else $L$ \fi}
\def  \SpecLum    {\ifmmode \Lunimosity_\nu \else $\Lunimosity_\nu$ \fi} 
\def  \Lsun       {\ifmmode \Lunimosity_\sun \else $\Lunimosity_\sun$ \fi}
\def  \Lsunh      {\ifmmode \h^{-2}\,\Lsun\else    $\h^{-2}\,\Lsun$\fi}      

\def \B           {\mbox{B}}   
\def \V           {\mbox{V}}   
\def \R           {\mbox{R}}   
\def \NUV         {\mbox{NUV}} 
\def \Ks          {\ifmmode \mbox{K}_{\mbox{\scriptsize s}}\else $\mbox{K}_{\mbox{\scriptsize s}}$\fi} 

\def \m           {\mbox{m}} 
\def \SDSSu       {\ifmmode u\else $u$\fi} 
\def \SDSSg       {\ifmmode g\else $g$\fi} 
\def \SDSSr       {\ifmmode r\else $r$\fi} 
\def \SDSSi       {\ifmmode i\else $i$\fi} 
\def \SDSSz       {\ifmmode z\else $z$\fi} 
\def \magFUV      {\ifmmode \m_{\mbox{\scriptsize FUV}}\else $\mbox{m}_{\mbox{\scriptsize FUV}}$\fi} 
\def \magNUV      {\ifmmode \m_{\mbox{\scriptsize NUV}}\else $\mbox{m}_{\mbox{\scriptsize NUV}}$\fi} 
\def \magKs       {\ifmmode \mbox{K}_{\mbox{\scriptsize s}}\else $\mbox{K}_{\mbox{\scriptsize s}}$\fi} 
\def \magWThree   {\ifmmode \m_{\mbox{\scriptsize W3}}\else $\mbox{m}_{\mbox{\scriptsize W3}}$\fi} 
\def \magWFour    {\ifmmode \m_{\mbox{\scriptsize W4}}\else $\mbox{m}_{\mbox{\scriptsize W4}}$\fi} 

\def \AbsMFUV    {\ifmmode \mbox{M}_{\mbox{\scriptsize FUV}}\else $\mbox{M}_{\mbox{\scriptsize FUV}}$\fi} 
\def \AbsMNUV    {\ifmmode \mbox{M}_{\mbox{\scriptsize NUV}}\else $\mbox{M}_{\mbox{\scriptsize NUV}}$\fi} 
\def \AbsMJ      {\ifmmode \mbox{M}_{\mbox{\scriptsize J}}\else $\mbox{M}_{\mbox{\scriptsize J}}$\fi}  
\def \AbsMH      {\ifmmode \mbox{M}_{\mbox{\scriptsize H}}\else $\mbox{M}_{\mbox{\scriptsize H}}$\fi}  
\def \AbsMKs     {\ifmmode \mbox{M}_{\mbox{\scriptsize K_s}}\else $\mbox{M}_{\mbox{\scriptsize K_s}}$\fi} 
\def \AbsMWThree {\ifmmode \mbox{M}_{\mbox{\scriptsize W3}}\else $\mbox{M}_{\mbox{\scriptsize W3}}$\fi} 
\def \AbsMWFour  {\ifmmode \mbox{M}_{\mbox{\scriptsize W4}}\else $\mbox{M}_{\mbox{\scriptsize W4}}$\fi} 
\def \AbsMg       {\ifmmode \mbox{M}_{g}\else $\mbox{M}_{g}$\fi} 

\def \AbsMgMax    {\ifmmode \mbox{M}_{g,max}\else $\mbox{M}_{g,max}$\fi} 
\def \AbsMgSun    {\ifmmode \mbox{M}_{g,\sun}\else $\mbox{M}_{g,\sun}$\fi} 

\def \magAsecSq {\ifmmode \mbox{mag\,arcsec}^{-2}\else  $\mbox{mag\,arcsec}^{-2}$\fi} 

\def \BMinR       {\ifmmode \mbox{{\B}-{\R}}\else {\B}-{\R}\fi} 
\def \BMinV       {\ifmmode \mbox{{\B}-{\V}}\else {\B}-{\V}\fi} 
\def \gMinr       {\ifmmode \mbox{{\SDSSg}-{\SDSSr}}\else {\SDSSg}-{\SDSSr}\fi} 
\def \gMini       {\ifmmode \mbox{{\SDSSg}-{\SDSSi}}\else {\SDSSg}-{\SDSSi}\fi} 
\def \uMinr       {\ifmmode \mbox{{\SDSSu}-{\SDSSr}}\else {\SDSSu}-{\SDSSr}\fi} 
\def \NUVMinu     {\ifmmode \mbox{{\NUV}-{\SDSSu}}\else {\NUV}-{\SDSSu}\fi} 
\def \NUVMinr     {\ifmmode \mbox{{\NUV}-{\SDSSr}}\else {\NUV}-{\SDSSr}\fi} 

\def \EBV         {\ifmmode \mbox{E}_{\mbox{\scriptsize \BMinV}}\else $\mbox{E}_{\mbox{\scriptsize \BMinV}}$\fi} 

\def \F          {\mbox{F}} 

\def \Whalf      {\ifmmode W_{50}\else $W_{50}$\fi}          
\def \FHI        {\ifmmode \mbox{F}_{HI} \else $\F_{HI}$\fi} 
\def \logMHI     {\ifmmode \log\left(\MHI/\Msun\right)\else $\log\left(\MHI/\Msun\right)$\fi } 

\def  \ergPerS     {\ifmmode {\rm ergs\,s}^{-1} \else ergs s$^{-1}$ \fi}
\def  \ergcms      {\ifmmode {\rm ergs\,cm}^{-2}\,{\rm s}^{-1} 
                       \else ergs\,cm$^{-2}$\,s$^{-1}$ \fi}
\def  \ergcmsA     {\ifmmode{\rm ergs\,cm}^{-2}\,{\rm s}^{-1}\,{\rm\AA}^{-1}
                       \else ergs\,cm$^{-2}$\,s$^{-1}$\,\AA$^{-1}$ \fi}
\def  \ergcmsHz    {\ifmmode{\rm ergs\,cm}^{-2}\,{\rm s}^{-1}\,{\rm Hz}^{-1}
                       \else ergs\,cm$^{-2}$\,s$^{-1}$\,Hz$^{-1}$ \fi}

\def  \Metallicity {\ifmmode Z\else $Z$\fi}
\def  \Zsun        {\ifmmode \Metallicity_\sun\else $\Metallicity_\sun$\fi}

\def \SFR       {\ifmmode \mbox{SFR}\else $\mbox{SFR}$\fi} 
\def \SSFR      {\ifmmode \mbox{SSFR}\else $\mbox{SSFR}$\fi} 
\def \SFE       {\ifmmode \mbox{SFE}\else $\mbox{SFE}$\fi} 
\def \MsunPerYr {\ifmmode \Msun\,\yr^{-1}\else $\Msun\,yr^{-1}$\fi} 

\def  \Halpha   {\ifmmode {{\rm H}_\alpha}\else {H$_\alpha$}\fi}  
\def  \nHa      {\ifmmode {{\rm nH}_\alpha}\else {nH$_\alpha$}\fi}  
\def  \Hbeta    {\ifmmode {\rm H}_\beta \else H$_\beta$ \fi}
\def  \Hgamma   {\ifmmode {\rm H}_\gamma \else H$_\gamma$ \fi}
\def  \Hdelta   {\ifmmode {\rm H}_\delta \else H$_\delta$ \fi}
\def  \Lya      {\ifmmode {\rm Ly}_\alpha \else Ly$_\alpha$ \fi}
\def  \Lyb      {\ifmmode {\rm Ly}_\beta \else Ly$_\beta$ \fi}
\def  \Pa       {\ifmmode {\rm P}_\alpha \else P$_\alpha$ \fi}
\def  \Pb       {\ifmmode {\rm P}_\beta \else P$_\beta$ \fi}
\def  \Bra      {\ifmmode {\rm Br}_\alpha \else Br$_\alpha$ \fi}
\def  \Brg      {\ifmmode {\rm Br}_\gamma \else Br$_\gamma$ \fi}

%% file: Chapters/Abstract.tex
\begin{abstract}

We have selected a sample of extremely isolated galaxies (EIGs) from the local Universe ($\z < 0.024$), using a simple isolation criterion: having no known neighbours closer than 300\,\kms ({3\,\Mpch}) in the three-dimensional redshift space $(\alpha,\delta,\z)$. The sample is unique both in its level of isolation and in the fact that it utilizes HI redshifts from the Arecibo Legacy Fast ALFA survey (ALFALFA).
We analysed the EIG sample using cosmological simulations and found that it contains extremely isolated galaxies with normal mass haloes which have evolved gradually with little or no ``major events'' (major mergers, or major mass-loss events) in the last {3\,\Gyr}. The fraction of EIGs which deviate from this definition (false positives) is 5\%--10\%.
For the general population of dark matter haloes it was further found that the mass accretion (relative to the current halo mass) is affected by the halo environment mainly through strong interactions with its neighbours. As long as a halo does not experience major events, its Mass Accretion History (MAH) does not depend significantly on its environment.
``Major events'' seem to be the main mechanism that creates low-mass subhaloes ($\Mhalo < 10^{10}\,\Msunh$) that host galaxies (with $\AbsMg \lesssim -14$).

\end{abstract}

%% file: Chapters/1_Introduction.tex
\section{Introduction and Background}
\label{ch:Introduction}

The research described here is part of an extensive study of star formation properties and evolution of galaxies in different environments and of various morphological types, conducted in the past few decades \citep[e.g.,][]{1983PhDT.........1B, 1995PhDT........86A, 1998MNRAS.298..920A, 1998ApJ...504..720B, 2001PhDT..Ana_Heller, 2003PhDT..Carmia_Weingarten, 2008arXiv0806.2722B, 2008MNRAS.390..408Z}. Specifically, we studied galaxies in the most extremely underdense regions of the local Universe. These galaxies are particularly interesting since they evolved with little or no environmental interference, and are therefore useful for validating and calibrating galaxy evolution models. Furthermore, when compared to galaxies in denser regions, they illuminate the overall effects of the environment on the evolution of galaxies.

We have chosen a sample of extremely isolated galaxies (EIGs) from the local Universe ($\z < 0.024$), based on a simple isolation criterion. The neighbourhood properties of this sample were analysed in detail using both observational data and cosmological simulations. The cosmological simulations were further used to estimate the properties and histories of the dark matter (DM) haloes in which the sample EIGs reside.

One of the unique advantages of the EIG sample we study here is that, apart from the optical redshift data commonly used to estimate environment density, it also utilized HI redshifts from the Arecibo Legacy Fast ALFA survey (ALFALFA; \citealt{2011AJ....142..170H}). 
The ALFALFA survey is a second-generation untargeted extragalactic HI survey initiated in 2005 \citep{2005AJ....130.2598G, 2007AJ....133.2569G, 2007AJ....133.2087S}. This survey utilizes the superior sensitivity and angular resolution of the Arecibo 305\,m radio telescope to conduct the deepest ever census of the local HI Universe.
\rem{This ensures higher sensitivity to late-type galaxies with low luminosity and large gas content in comparison with other samples of isolated galaxies. }ALFALFA was particularly useful in verifying the isolation of the target galaxies, since by being an HI survey it easily measures redshifts of low surface brightness galaxies (LSBs) and other low-luminosity late-type neighbours that are often difficult to detect optically but abound with HI\rem{\citep{2013JApA...34...19D}}.

When analyzing the neighbourhood content, one should not ignore the possible presence of large invisible masses near the target galaxy. These ``dark haloes'' or ``dark galaxies'' can be composed of dark as well as non-luminous baryonic matter. Such dark galaxies with masses up to $10^{11}\,\Msun$ might exist \citep{2005ApJ...618..214T}, and their number density may be comparable to or even exceed by an order of magnitude that of luminous galaxies \citep{1999ApJ...522...82K}.
\cite{2006A&A...451..817K} estimated that the density of dark galaxies is less than {$\sim$1/20} of the population of luminous galaxies. This estimate came from a search for effects of interactions with neighbouring galaxies in some 1,500 isolated galaxies. They found that no more than 0.3\% of the isolated galaxies were disturbed to a noticeable level by dark galaxies.
If a dark galaxy contains sufficient HI, it might be detected \markChange{by} ALFALFA. In the latest ALFALFA catalogue \citep[$\alpha$.40;][]{2011AJ....142..170H} 199 such dark galaxies were found.  Many of these are suspected to be tidal or ram-pressure debris of nearby galaxies \citep[e.g.,][]{2008ApJ...682L..85K}.
ALFALFA is, therefore, an extremely important tool for testing the isolation of galaxies.

Extensive optical imaging of the sample EIGs in broad-band and rest-frame {\Halpha} was performed using the Wise Observatory\footnote{http://wise-obs.tau.ac.il/} (WO) one meter telescope. This, along with public observational data, were used to measure the current star formation (following the method described in \citealt{2012MNRAS.419.2156S}) and to estimate its history. These observational results will be described and discussed in detail in Spector \& Brosch (in preparation). 

This work attempts, among other things, to help resolve the question of ``Nature vs. Nurture''; does the evolution of galaxies depend only on their content or do their large-scale environments have a significant evolutionary influence.
Some argue that galaxy formation is driven predominantly by the mass of the host DM halo, and is nearly independent of the larger-scale halo environment (e.g., \citealt{2008MNRAS.386.2285C, 2009ApJ...691..633T}). This is supported by their simulation models that produce void galaxies conforming to some observed statistical properties. However, since there are many galaxy properties that current simulations cannot predict, and since the halo mass of galaxies cannot be directly measured, this hypothesis is hard to prove or disprove.

For similar purposes, other samples of isolated galaxies were defined and studied 
in ``the Analysis of the interstellar Medium of Isolated GAlaxies'' (AMIGA) international project \citep{2007A&A...472..121V, 2013MNRAS.434..325F}\rem{http://amiga.iaa.es}, 
in the ``Two Micron Isolated Galaxy'' catalogue (2MIG; \citealt{2010AstBu..65....1K}), 
in the ``Local Orphan Galaxies'' catalogue (LOG; \citealt{2011AstBu..66....1K, 2013AstBu..68..243K}), 
and in the Void Galaxy Survey (VGS;  \citealt{2012AJ....144...16K}).
These are discussed in section \ref{s:Envrnmnt_CompOther}.

\vspace{12pt}

In section \ref{ch:Sample} the method used for selecting the sample of extremely isolated galaxies (EIGs) is described, the sample galaxies are listed, and their observed neighbourhoods are discussed and compared to those of other isolated galaxy surveys.
Section \ref{s:Envrnmnt_Simulations} describes how cosmological simulations were used to analyse the EIG sample, and discusses the estimated halo properties and histories of the EIGs, as well as the properties of their neighbourhoods.

\vspace{12pt}

Throughout this work, unless indicated otherwise, Lambda Cold Dark Matter ($\Lambda$CDM) cosmology with the seven-year Wilkinson Microwave Anisotropy Probe data (WMAP7, \citealt{2011ApJS..192...17B}) parameters are used, including the dimensionless Hubble parameter $\h = 0.704$.
We adopt here the solar {\SDSSg}-band absolute magnitude of $\AbsMgSun = +5.12$ (according to the Sloan Digital Sky Survey, SDSS, DR7 web site\footnote{www.sdss.org/dr7/algorithms/sdssUBVRITransform.html\-\#vega\_sun\_colors}).

%% file: Chapters/2_TheSample.tex
\section{The Sample}
\label{ch:Sample}

The first and possibly most crucial step in this study is selecting a sample of galaxies in extremely underdense regions, referred to here as Extremely Isolated Galaxies (EIGs). By definition, these EIGs are very rare and, therefore, at moderate redshifts the sample is expected to be fairly small, including only a few dozen galaxies. This section describes the method used for selecting the sample, lists the sample galaxies, discusses their observed neighbourhoods, and compares these with other isolated galaxy catalogues.

\rem{
The ALFALFA HI survey \citep{2005AJ....130.2598G, 2011AJ....142..170H}, which covers $\sim7,000\,\sqDeg$ of the sky with faint HI detection limit and excellent positioning accuracy, is a key data source for selecting the sample and analyzing its HI content. Further observational data are obtained from NED and SDSS \citep{2014ApJS..211...17A\rem{SDSS DR10}}. 
}

\subsection{Isolation criterion}
\label{sec:Sample.Criterion}

In the last few decades great advances were made in redshift surveys, which now map the local Universe in redshift space with great precision. Before these became available, isolated galaxies (IGs) had to be identified using projected coordinates alone, i.e.~searching in two-dimensional space (2D). Radial distances had to be estimated, for example, using the angular sizes of galaxies, such as done for the classical Catalogue of Isolated Galaxies \citep[CIG;][]{1973SoSAO...8....3K} and in \cite{2010AstBu..65....1K}.

The use of redshift data for testing the isolation of galaxies started decades ago \citep{1977ApJ...216..694H}. Nowadays, when the local Universe is mapped in detail, it is possible to perform accurate three-dimensional (3D) redshift space searches. The advantages of using such strategy are simplicity and straightforwardness, not having to assume anything about the characteristics of the galaxies (size, magnitude, etc.).

However, using redshift mapping introduces two difficulties. First is the incompleteness of most redshift databases. A galaxy that seems to be isolated might have neighbours for which a redshift was not yet measured. Second is the error in radial distance introduced by peculiar velocities. Using redshift data one performs a search in 3D redshift space, the mathematical representation of the projected coordinates: right ascension ({\RA}) and declination ({\Dec}), and the radial coordinate: redshift (\z). It should be kept in mind that mapping in 3D redshift space ({\RA}, {\Dec}, \z) can differ significantly from the true mapping in real space\rem{({\RA}, {\Dec}, r)}. For example, although close in real space, two galaxies in a cluster might have very different redshifts due to the cluster velocity dispersion and will therefore seem distant in 3D redshift space.

In principle, independent distance measurements (not based on redshift measurements) have the potential to improve uncertainties caused by peculiar velocities. Currently, independent distance measurements are available for more than 8000 galaxies in the local Universe. These were used by \cite{2014Natur.513...71T} to find the limits of the ``Laniakea'' super-cluster in which we live.
The accuracy of these distance measurements for very close galaxies and for early type galaxies (typically located in dense regions) is $\sim$10\%, while the accuracy for the more distant late types (typically located in isolated regions) is $\sim$20\% \citep{2012ApJ...744...43C}. When averaged, these give sufficient accuracy to map cosmic flows. However, for testing the isolation of individual galaxies in the redshift range of this research, these independent distance measurements are not accurate enough.

\vspace{12pt}

In this work we have chosen to use the simple isolation criterion described in \cite{2010ASPC..421...27S}. {\bf A galaxy is considered an EIG and is included in the sample if it has no known neighbours closer than {300\,\kms} in 3D redshift space, and if its redshift is in the range $2000<\cz<7000\,\kms$}. This translates to not having any known neighbour within a distance of $3\,\Mpch \cong 4.26\,\Mpc$.\footnote{This criterion describes the most isolated subsample of EIGs studied here. Not all sample galaxies (EIGs) pass this isolation criterion (see section \ref{sec:Sample.TheIsoGals}).}

The redshift range was limited to {7000\,\kms} to have reasonable completeness of redshift data around each galaxy. The reason for the lower limit of {2000\,\kms} is to keep the sky area that has to be searched around each galaxy relatively small, since at {2000\,\kms} neighbours have to be searched for as far as 8.6$\deg$ away.

No magnitude, HI mass or size limit was used in the selection of candidate neighbours. The use of such limits would have somewhat reduced the level of isolation of the sample (especially for the closer EIGs), and therefore was not preferred. Not using such limits, however, complicates somewhat the analysis of the sample's isolation level (described in section \ref{s:Envrnmnt_Simulations}).

\subsection{Selection process}
\label{s:SampleProcess}

The search criterion was applied to two sky regions, one in the spring sky (Spring) and the other in the autumn sky (Autumn), as described in Table \ref{T:Sample-Regions}. These particular regions were selected since they are covered by the $\alpha$.40 ALFALFA catalogue \citep{2011AJ....142..170H}. Both regions include mainly high Galactic latitudes. The Spring region is almost fully covered by spectroscopic data in SDSS DR10 \citep{2014ApJS..211...17A}.

\begin{ctable} 
[
  caption = Sample search regions,
  doinside = \small,
  star,
  label   = T:Sample-Regions
]
{@{}ccccc@{}}
{}
{
  \FL
          & {\RA} (J2000)      & {\Dec} (J2000)         & \cz~$\left[ \kms \right]$  & Volume $\left[ \VolMpch \right]$
  \ML
  Spring  & 7h30m--16h30m   & $4\deg$--$16\deg$   & 2000--7000  & $5.42 \cdot 10^{4}$
  \NN
  Autumn  & 22h00m--03h00m  & $24\deg$--$28\deg$  & 2000--7000  & $9.17 \cdot 10^{3}$
  \LL
}

\end{ctable}

The Spring region contains parts of the following large-scale structures (ordered by increasing redshift): \markChange{Virgo cluster}, Virgo void, Coma void, Microscopium void, and the Coma wall. The Autumn region contains parts of the following large-scale structures (ordered by increasing redshift): Delphinus void, Taurus void, Eridanus void, Perseus Pisces supercluster, Pegasus void, and Pisces void \citep{1998Book..Fairall}.

To allow searching for neighbours near the edges of the these search regions, larger database regions were used. The region downloaded from the NASA/IPAC Extragalactic Database\footnote{http://ned.ipac.caltech.edu/} (NED) for the Spring region was: $\mbox{6h40m} < {\RA} < \mbox{17h20m}$, ${-8\deg} < {\Dec} < {28\deg}$, $1600 < {\cz} < 7400\,\kms$. The region downloaded from NED for the Autumn region was: $\mbox{21h00m} < {\RA} < \mbox{04h00m}$, ${12\deg} < {\Dec} < {40\deg}$, $1600 < {\cz} < 7400\,\kms$. These regions allow searching for neighbours at distances of up to {400\,\kms} (equivalent to {4\,\Mpch}) from the candidate galaxies.

Data for these regions were downloaded from NED on November 13, 2012. The NED object types included in the database were: galaxies, galaxy clusters, galaxy pairs, galaxy triples, galaxy groups, and QSO. The Spring database region included 14273 objects, while the Autumn database region included 3956 objects.

The ALFALFA database used was the ``$\alpha$.40 HI source catalogue'' \citep[$\alpha$.40;][]{2011AJ....142..170H}. This catalogue covers 40\% of the final ALFALFA survey area ($\sim$2800\,\sqDeg) and contains 15855 sources. It includes parts of the required Spring region: $\mbox{07h30m} < {\RA} < \mbox{16h30m}$, ${+04\deg} < {\Dec} < {+16\deg}$ and ${+24\deg} < {\Dec} < {+28\deg}$, and parts of the required Autumn region: $\mbox{22h} < {\RA} < \mbox{03h}$, ${+14\deg} < {\Dec} < {+16\deg}$ and ${+24\deg} < {\Dec} < {+32\deg}$. The database covers the required redshift range ($1600 < {\cz} < 7400\,\kms$). 

For each EIG found in these searches the NED redshift measurement was verified by comparing it to ALFALFA, SDSS DR10, and all sources quoted by NED. The redshift values adopted here for the EIGs were chosen based on the following priority list:
\begin{enumerate}
  \item If an optically-derived redshift value with uncertainty $<10\,\kms$ exists, it was adopted. If several such values exist, the SDSS value of the latest available data release (usually DR10) was preferred.
  \item Otherwise, if a 21cm redshift value with uncertainty $<10\,\kms$ exists, it was adopted. If several such values exist, the ALFALFA value was preferred.
  \item Otherwise, if reasonably accurate redshift values exist ($\Delta \cz < 30\,\kms$), the most accurate of them was adopted.
  \item If no reasonably accurate value exists (i.e., $\Delta \cz \geq 30\,\kms$), the galaxy was deleted from the sample.
  \item[]
\end{enumerate}

Optical redshift measurements were preferred because they are expected to be more accurate for estimating the transmittance of the {\Halpha} filters in the redshifted {\Halpha} line.

The EIGs' isolation was tested again, using the adopted redshift. The neighbourhoods of all EIGs were then evaluated using data downloaded from NED for spheres of {10\,\Mpch} around each EIG.

\subsection{The Extremely Isolated Galaxies (EIGs)}
\label{sec:Sample.TheIsoGals}

Only 14 galaxies were found to be isolated, according to the search criterion stated above, in the Spring region. This corresponds to a fraction of $\left(0.4^{+0.3}_{-0.2} \right)\%$ of the galaxies in the Spring NED dataset. In the Autumn sky region, 6 galaxies were found to be isolated according to the criterion. This corresopnds to $\left(1.5^{+1.8}_{-0.8} \right)\%$ of the galaxies in the Autumn NED dataset. An additional Autumn galaxy, EIG 1a-04, was added to the sample although it lies outside the search region.

The larger fraction of EIGs in the Autumn region can be attributed to the fact that the Spring region is fully covered by SDSS (with spectroscopic data), while the Autumn region is not. It may also be attributed to the different composition of both regions, where the Spring region may contain a significantly larger fraction of cluster and wall galaxies. 

The use of the ALFALFA unbiased HI data significantly improved the quality of the sample. \markChange{Out of 32 galaxies that passed the criterion using NED data alone, 11 galaxies did not pass the criterion when tested with ALFALFA data (seven in the Spring region}, and four in the Autumn region). \markChange{For the seven Spring region galaxies 13 neighbours were found in ALFALFA}, and for the four Autumn region galaxies 10 neighbours were found.

The galaxies studied here were divided to three subsamples:
\begin{enumerate}
  \item[1.] Galaxies that passed the criterion using both NED and ALFALFA data.
  \item[2.] Galaxies that passed the criterion using NED data, but did not pass using ALFALFA data (had neighbours closer than {3\,\Mpch} in the ALFALFA database).
  \item[3.] Galaxies for which the distance to the closest neighbour in NED's data is {2 -- 3\,\Mpch} (regardless of the distance to the closest neighbour in ALFALFA's data).
  \item[]
\end{enumerate}

Subsamples 1 and 2 are complete, in the sense that they contain all catalogued galaxies that passed their criteria in the studied sky regions. Subsample 3 is far from being complete. It contains only those galaxies that seemed to be isolated in the present or earlier searches, but were later found to have neighbours in the range {2 -- 3\,\Mpch} (\markChange{7 in the Spring region}, and 2 in the Autumn region). These include galaxies for which neighbours were added to NED or ALFALFA in recent years, as well as galaxies which had a low-accuracy redshift value in NED and for which using a more accurate redshift value yielded closer neighbours. It also contains a galaxy, EIG 3s-06, which was found by searching the ALFALFA data alone, but had neighbours in the range {2 -- 3\,\Mpch} in the NED dataset.

The galaxies were named according to their subsample and sky region, using the following format:
\begin{equation*}
\mbox{EIG BR-XX}
\end{equation*}
where:
\begin{description}
  \item[B] is the galaxy's subsample (1, 2 or 3, as described above);
  \item[R] is the sky region (``s'' - Spring, ``a'' - Autumn);
  \item[XX] is the serial number of the galaxy in the subsample.
  \item[]
\end{description}

So, for example, object EIG 3s-06 is the sixth galaxy in subsample 3 of the spring sky region.

\vspace{12pt}

The galaxies of the different subsamples are listed in Tables \ref{T:EIG-1s} through \ref{T:EIG-3a}. The data for each galaxy include its EIG name, the first name listed for it in NED, its ALFALFA name, and its coordinates in redshift space ({\RA}, {\Dec}, {\cz}).

\begin{ctable} 
[
  cap     = {The EIG-1s subsample},
  caption = {The EIG-1s subsample - Spring region galaxies with no neighbours closer than {3\,\Mpch} in both NED and ALFALFA data},
  doinside = \small,
  star,
  label   = {T:EIG-1s}
]
{@{}cp{4.7cm}cccr@{$\ \pm\:$}r@{}}
{}
{
  \FL
  Name       & NED ID                      & ALFALFA ID        & {\RA}      & {\Dec}    & \multicolumn {2}{c}{\cz}
  \NN
             &                             &                   & (J2000)    & (J2000)   & \multicolumn {2}{c}{$\left[ \kms \right]$}
  \ML
  EIG  1s-01 & SDSS  J075041.99+144717.3   & HI075041.7+144741 & 07:50:42.0 & +14:47:17 & 5399&3
  \NN
  EIG  1s-02 & 2MASX  J08061617+1249401    & HI080614.1+125021 & 08:06:16.1 & +12:49:41 & 5694&2
  \NN
  EIG  1s-03 & UGC  04655                  & HI085333.4+044710 & 08:53:32.7 & +04:46:57 & 6189&1
  \NN
  EIG  1s-04 & SDSS  J092131.91+112048.2   & HI092131.3+112100 & 09:21:31.9 & +11:20:48 & 5670&7
  \NN
  EIG  1s-05 & AGC  208312                 & HI102039.6+080914 & 10:20:39.6 & +08:09:06 & 5336&5
  \NN
  EIG  1s-06 & SDSS  J102352.85+062417.0   & HI102352.7+062416 & 10:23:52.8 & +06:24:17 & 5587&3
  \NN
  EIG  1s-07 & SDSS  J110414.59+050736.6   & HI110418.1+050703 & 11:04:14.6 & +05:07:37 & 5269&1
  \NN
  EIG  1s-08 & SDSS  J111624.13+054352.7   & -                 & 11:16:24.1 & +05:43:53 & 4976&1
  \NN
  EIG  1s-09 & SDSS  J112156.76+102955.3   & HI112157.6+102948 & 11:21:56.8 & +10:29:55 & 4453&2
  \NN
  EIG  1s-10 & SDSS  J124011.52+154213.8   & HI124009.5+154213 & 12:40:11.5 & +15:42:14 & 3916&1
  \NN
  EIG  1s-11 & VCC  1889                   & -                 & 12:41:46.1 & +11:15:02 & 4725&10
  \NN
  EIG  1s-12 & SDSS  J133156.93+133101.6   & -                 & 13:31:56.9 & +13:31:02 & 4864&1
  \NN
  EIG  1s-13 & SDSS  J151410.95+064449.0   & -                 & 15:14:10.9 & +06:44:49 & 5427&2
  \NN
  EIG  1s-14 & CGCG  050-112               & HI155029.2+042810 & 15:50:25.5 & +04:28:35 & 6122&17
  \LL
}
\end{ctable}

\begin{ctable} 
[
  cap     = {The EIG-1a subsample},
  caption = {The EIG-1a subsample - Autumn region galaxies with no neighbours closer than {3\,\Mpch} in both NED and ALFALFA data},
  doinside = \small,
  star,
  label   = {T:EIG-1a}
]
{@{}cp{4.7cm}cccr@{$\ \pm\:$}r@{}}
{}
{
  \FL
  Name       & NED ID                      & ALFALFA ID        & {\RA}      & {\Dec}    & \multicolumn {2}{c}{\cz}
  \NN
             &                             &                   & (J2000)    & (J2000)   & \multicolumn {2}{c}{$\left[ \kms \right]$}
  \ML
  EIG  1a-01 & 2MASX  J00270759+2459072   & HI002706.2+245912   & 00:27:07.6 & +24:59:07 &6378&12
  \NN
  EIG  1a-02 & 2MASX  J00563772+2418526   & HI005632.5+241856   & 00:56:37.7 & +24:18:53 &6501&18
  \NN
  EIG  1a-03 & AGC  122211   & HI023136.3+263250   & 02:31:36.8 & +26:32:30 &3691&1
  \NN
  EIG  1a-04 & IC  0238 & - & 02:35:22.7 & +12:50:16 &6008&21
  \NN
  EIG  1a-05 & 2MASX  J02535284+2630267   & HI025352.1+263035   & 02:53:52.9 & +26:30:27 &6176&3
  \NN
  EIG  1a-06 & AGES  J025917+244756   & - & 02:59:17.5 & +24:48:43 &4658&3
  \NN
  EIG  1a-07 & AGC  321304   & HI220351.1+252659   & 22:03:51.1 & +25:26:32 &2692&13
  \LL
}
\end{ctable}

\begin{ctable} 
[
  cap     = {The EIG-2s subsample},
  caption = {The EIG-2s subsample - Spring region galaxies with no neighbours closer than {3\,\Mpch} in NED data, but some in ALFALFA data},
  doinside = \small,
  star,
  label   = {T:EIG-2s}
]
{@{}cp{4.7cm}cccr@{$\ \pm\:$}r@{}}
{}
{
  \FL
  Name       & NED ID                      & ALFALFA ID        & {\RA}      & {\Dec}    & \multicolumn {2}{c}{\cz}
  \NN
             &                             &                   & (J2000)    & (J2000)   & \multicolumn {2}{c}{$\left[ \kms \right]$}
  \ML
  EIG  2s-01 & SDSS  J075532.17+113316.7   & -                 & 07:55:32.2 & +11:33:17 &5842&3
  \NN
  EIG  2s-02 & LSBC  F704-V01              & HI082452.4+091319 & 08:24:51.7 & +09:13:29 &6018&2
  \NN
  EIG  2s-04 & SDSS  J124548.06+092029.0   & HI124548.6+092025 & 12:45:48.0 & +09:20:29 &5740&4
  \NN
  EIG  2s-05 & CGCG  076-069               & HI144932.9+134845 & 14:49:33.8 & +13:48:25 &5647&5
  \NN
  EIG  2s-06 & CGCG  050-028               & HI153445.2+061813 & 15:34:46.1 & +06:17:53 &6313&1
  \NN
  EIG  2s-07 & SDSS  J154627.10+083924.8   & -                 & 15:46:27.1 & +08:39:25 &3711&1
  \NN
  EIG  2s-08 & SDSS  J161517.02+130133.0   & -                 & 16:15:17.0 & +13:01:33 &3650&1
  \LL
}
\end{ctable}

\begin{ctable} 
[
  cap     = {The EIG-2a subsample},
  caption = {The EIG-2a subsample - Autumn region galaxies with no neighbours closer than {3\,\Mpch} in NED data, but some in ALFALFA data},
  doinside = \small,
  star,
  label   = {T:EIG-2a}
]
{@{}cp{4.7cm}cccr@{$\ \pm\:$}r@{}}
{}
{
  \FL
  Name       & NED ID                      & ALFALFA ID        & {\RA}      & {\Dec}    & \multicolumn {2}{c}{\cz}
  \NN
             &                             &                   & (J2000)    & (J2000)   & \multicolumn {2}{c}{$\left[ \kms \right]$}
  \ML
  EIG  2a-01 & CGCG  480-041   & HI010617.0+253240   & 01:06:11.9 & +25:33:06 &6623&9
  \NN
  EIG  2a-02 & FGC  0362   & HI025608.0+274210   & 02:56:08.6 & +27:42:02 &6473&8
  \NN
  EIG  2a-03 & KUG  2239+275   & HI224205.5+274630   & 22:42:07.3 & +27:46:11 &6964&2
  \NN
  EIG  2a-04 & AGC  321226   & HI225542.2+261830   & 22:55:44.8 & +26:18:10 &4372&12
  \LL
}
\end{ctable}

\begin{ctable} 
[
  cap     = {The EIG-3s subsample},
  caption = {The EIG-3s subsample - Spring region galaxies, for which the closest neighbour in NED data is  at a distance of {2 -- 3\,\Mpch}},
  doinside = \small,
  star,
  label   = {T:EIG-3s}
]
{@{}cp{4.7cm}cccr@{$\ \pm\:$}r@{}}
{}
{
  \FL
  Name       & NED ID                      & ALFALFA ID        & {\RA}      & {\Dec}    & \multicolumn {2}{c}{\cz}
  \NN
             &                             &                   & (J2000)    & (J2000)   & \multicolumn {2}{c}{$\left[ \kms \right]$}
  \ML
  EIG  3s-01 & SDSS  J104008.81+091628.5   & HI104008.7+091607 & 10:40:08.8 & +09:16:29 &5420&3
  \NN
  EIG  3s-02 & SDSS  J123814.44+100949.8   & HI123813.8+100902 & 12:38:14.4 & +10:09:50 &5840&11
  \NN
  EIG  3s-03 & CGCG  043-046 & HI125133.7+080242 & 12:51:33.5 & +08:02:43 &3620&2
  \NN
  EIG  3s-04 & AGC  225879   & HI125829.0+121115 & 12:58:30.5 & +12:11:22 &4085&5
  \NN
  EIG  3s-05 & CGCG  047-124   & HI143846.4+073700 & 14:38:46.8 & +07:37:03 &5527&3
  \NN
  EIG  3s-06 & SDSS  J150544.49+111230.1   & HI150544.8+111203 & 15:05:44.5 & +11:12:30 &3545&2
  \NN
  EIG  3s-07 & SDSS  J151054.61+054314.7   & HI151055.9+054325 & 15:10:54.6 & +05:43:15 &6436&4
  \LL
}
\end{ctable}

\begin{ctable} 
[
  cap     = {The EIG-3a subsample},
  caption = {The EIG-3a subsample - Autumn region galaxies, for which the closest neighbour in NED data is  at a distance of {2 -- 3\,\Mpch}},
  doinside = \small,
  star,
  label   = {T:EIG-3a}
]
{@{}cp{4.7cm}cccr@{$\ \pm\:$}r@{}}
{}
{
  \FL
  Name       & NED ID                      & ALFALFA ID        & {\RA}      & {\Dec}      & \multicolumn {2}{c}{\cz}
  \NN
             &                             &                   & (J2000)    & (J2000)   & \multicolumn {2}{c}{$\left[ \kms \right]$}
  \ML
  EIG  3a-01 & UGC  12123   & HI223752.8+251146   & 22:37:53.4 & +25:11:36 &4082&2
  \NN
  EIG  3a-02 & 2MASX  J01331560+2614556   & HI013314.5+261508   & 01:33:15.6 & +26:14:55 &6952&1
  \LL
}
\end{ctable}

\markChange
{Notes regarding specific EIGs are listed in appendix \ref{App:EIGdata}. Section \ref{App:EIGdata.Deleted} of this appendix lists the objects that were first found to be isolated, but were eventually not included in the sample for the various reasons described there.
}

\subsection{Observed neighbourhoods}
\label{s:Envrnmnt_observed}

\begin{figure}
\begin{centering}
\includegraphics[width=8cm,trim=0mm 0mm 0mm 0, clip]{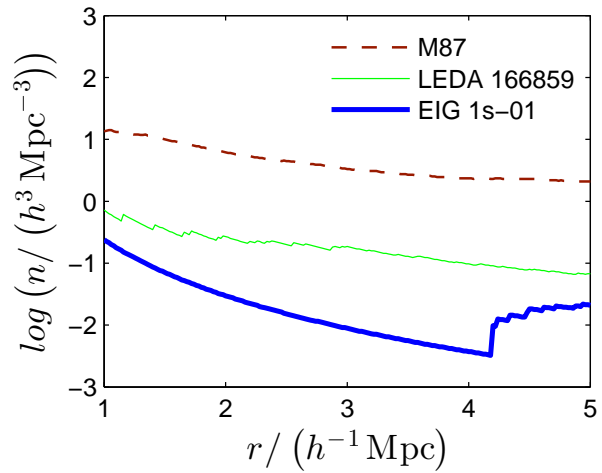}

\caption[Number density, $n$, in a sphere of varying radius, $r$, around three galaxies (EIG, field and cluster galaxy).]
{ The number density, $n$, in a sphere of varying radius, $r$, around three galaxies:
  EIG 1s-01, LEDA 166859 -- a field galaxy, M87 -- cluster galaxy. \label{f:Sample-n_r_3Gals}
}
\end{centering}
\end{figure}

First, an example of the huge difference between the environments of EIGs, field galaxies and cluster galaxies is illustrated in Figure \ref{f:Sample-n_r_3Gals}. This figure shows the number density \markChange{of galaxies}, $n$, around EIG 1s-01, around a typical field galaxy (LEDA 166859) and around M87, a supergiant elliptical galaxy located near the centre of the Virgo cluster. For each of these three galaxies $n$ is shown as a function of $r$, the radius of a sphere around the galaxy, for which $n$ was calculated. The number density shown in this figure includes galaxies with known redshifts as well as the central galaxy itself, \markChange{and is calculated for redshift space, i.e. without compensating for peculiar velocities}.

EIG 1s-01 has no neighbours closer than $4\,\Mpch$ but has ten neighbours at a distance of {4 -- 5\,\Mpch}. As can be seen in Figure \ref{f:Sample-n_r_3Gals}, in the range of calculated sphere radius ($1 <  r < 5\,\Mpch$) the neighbourhood density of EIG 1s-01 is about one order of magnitude lower than that of the typical field galaxy (LEDA 166859) and about two orders of magnitude lower than that of the cluster galaxy (M87). This typical example indicates that the EIGs are extreme field galaxies, significantly more isolated than the average.

\vspace{12pt}

Next, specific neighbourhood data of each EIG are listed. Table \ref{T:EIG12_Ngbr3Mpch} summarizes information about the observed neighbourhood of the EIGs of subsamples EIG-1 and EIG-2 (objects that passed the isolation criterion). The distance to the nearest known neighbour, \dist{1}, obtained separately from the NED and $\alpha$.40 datasets, is listed. For ALFALFA, the number of known neighbours up to a distance of {3\,\Mpch} (neighbour count) is also listed (for NED it is zero by definition) along with the dataset coverage where ``Full'' indicates that the sphere of radius {3\,\Mpch} around the galaxy is fully covered by the $\alpha$.40 dataset and ``Partial'' indicates that only a part of this sphere is covered. The table also lists the name of the void in which the EIG is located (or the names of two adjacent voids, in case of a nameless void). The void names are as defined in the ``Atlas of nearby large-scale structures'' of \cite{1998Book..Fairall}.

Table \ref{T:EIG3_Ngbr3Mpch} summarizes the observed neighbourhood data for subsample EIG-3 (those galaxies which fell short of passing the isolation criterion, but were still studied). In addition to the fields listed in Table \ref{T:EIG12_Ngbr3Mpch}, the table lists the neighbour counts obtained from the NED dataset.

\begin{ctable} 
[
  cap     = {Observed neighbourhood of EIGs (subsamples 1 \& 2)},
  caption = {Observed neighbourhood of EIGs (subsamples 1 \& 2)},
  star,
  label   = {T:EIG12_Ngbr3Mpch}
]
{@{}cccccc@{}}
{
  \tnote[a]
  { 
    ``Neighbour count'' and ``Coverage'' refer to a sphere of radius {3\,\Mpch} around the EIG.
  }
  \tnote[b]
  {
    A void between the two voids, whose names are listed.
  }
}
{
 \FL
      & NED & \multicolumn  {3}{c}{ALFALFA} \NN
 EIG  & \dist{1}                & \dist{1}                & Neighbour        & Coverage \tmark[a] & Void \NN
 name & $\left[ \Mpch \right]$  & $\left[ \Mpch \right]$  & count \tmark[a] &                    & name \NN
 \cmidrule(r){1-1}\cmidrule(rl){2-2}\cmidrule(rl){3-5}\cmidrule(l){6-6} 
 1s-01 & 4.19 & 4.32 & 0 & Partial & Canis Major                         \NN
 1s-02 & 3.22 & 3.95 & 0 & Full    & Canis Major                          \NN
 1s-03 & 4.35 & 5.99 & 0 & Partial & Canis Major                         \NN
 1s-04 & 3.02 & 3.01 & 0 & Full    & Ursa Major -- Canis Major \tmark[b] \NN
 1s-05 & 3.01 & 3.05 & 0 & Full    & Ursa Major -- Hydra \tmark[b]       \NN
 1s-06 & 3.11 & 3.11 & 0 & Partial & Ursa Major -- Hydra \tmark[b]       \NN
 1s-07 & 3.27 & 5.07 & 0 & Partial & Ursa Major -- Hydra \tmark[b]       \NN
 1s-08 & 4.02 & 4.18 & 0 & Partial & Leo -- Hydra \tmark[b]              \NN
 1s-09 & 3.07 & 3.30 & 0 & Full    & Leo -- Hydra \tmark[b]              \NN
 1s-10 & 3.02 & 3.22 & 0 & Partial & Coma                                \NN
 1s-11 & 3.91 & 4.04 & 0 & Full    & Coma                                \NN
 1s-12 & 3.28 & 3.96 & 0 & Partial & Coma                                \NN
 1s-13 & 5.46 & 3.73 & 0 & Partial & Microscopium                        \NN
 1s-14 & 3.45 & 3.40 & 0 & Partial & Microscopium                        \NN
 \cmidrule(r){1-1}\cmidrule(rl){2-2}\cmidrule(rl){3-5}\cmidrule(l){6-6} 
 1a-01 & 3.20 & 3.69 & 0 & Partial & Pisces                              \NN 
 1a-02 & 3.28 & 3.33 & 0 & Partial & Pisces                              \NN
 1a-03 & 3.95 & 4.98 & 0 & Partial & Taurus                              \NN
 1a-04 & 3.35 & 3.21 & 0 & Partial & Pisces                              \NN
 1a-05 & 3.10 & 3.10 & 0 & Partial & Pisces                              \NN
 1a-06 & 4.49 & 4.49 & 0 & Partial & Taurus                              \NN
 1a-07 & 4.42 & 4.43 & 0 & Partial & Delphinus                           \NN
 \cmidrule(r){1-1}\cmidrule(rl){2-2}\cmidrule(rl){3-5}\cmidrule(l){6-6} 
 2s-01 & 3.24 & 1.07 & 2 & Full    & Canis Major                         \NN
 2s-02 & 3.61 & 0.66 & 1 & Full    & Canis Major                         \NN
 2s-04 & 3.00 & 1.90 & 2 & Full    & Coma                                \NN
 2s-05 & 3.32 & 0.94 & 3 & Partial & Microscopium                        \NN
 2s-06 & 3.31 & 1.57 & 3 & Partial & Microscopium                        \NN
 2s-07 & 3.43 & 2.37 & 1 & Full    & Virgo -- Microscopium \tmark[b]     \NN
 2s-08 & 3.21 & 1.71 & 1 & Partial & Microscopium                        \NN
 \cmidrule(r){1-1}\cmidrule(rl){2-2}\cmidrule(rl){3-5}\cmidrule(l){6-6} 
 2a-01 & 3.11 & 1.19 & 1 & Partial & Pisces                              \NN
 2a-02 & 3.30 & 1.63 & 4 & Partial & Pisces                              \NN
 2a-03 & 3.47 & 1.60 & 4 & Full    & Pegasus                             \NN
 2a-04 & 3.14 & 2.74 & 1 & Partial & Pegasus
 \LL
}
\end{ctable}

\begin{ctable} 
[
  cap     = {Observed neighbourhood of EIGs (subsample 3)},
  caption = {Observed neighbourhood of EIGs (subsample 3)},
  star,
  label   = {T:EIG3_Ngbr3Mpch}
]
{@{}ccccccc@{}}
{
  \tnote[a]
  { 
    ``Neighbour count'' and ``Coverage'' refer to a sphere of radius {3\,\Mpch} around the EIG.
  }
  \tnote[b]
  {
    A void between the two voids, whose names are listed.
  }
}
{
 \FL
      & \multicolumn  {2}{c}{NED} & \multicolumn  {3}{c}{ALFALFA} \NN
 EIG  & \dist{1}                & Neighbour        & \dist{1}                & Neighbour        & coverage \tmark[a] & Void  \NN
 name & $\left[ \Mpch \right]$  & count \tmark[a] & $\left[ \Mpch \right]$  & count \tmark[a] &                    & name  \NN
 \cmidrule(r){1-1}\cmidrule(rl){2-3}\cmidrule(rl){4-6}\cmidrule(l){7-7} 
 3s-01 & 2.86 & 1 & 3.14 & 0 & Full    & Ursa Major -- Hydra \tmark[b] \NN
 3s-02 & 2.29 & 1 & 1.96 & 3 & Full    & Coma                          \NN
 3s-03 & 2.87 & 1 & 0.97 & 4 & Partial & Coma                          \NN
 3s-04 & 2.98 & 1 & 2.98 & 1 & Partial & Coma                          \NN
 3s-05 & 2.83 & 1 & 5.12 & 0 & Full    & Microscopium                  \NN
 3s-06 & 2.44 & 4 & 3.71 & 0 & Partial & Virgo                         \NN
 3s-07 & 2.73 & 1 & 3.82 & 0 & Partial & Microscopium                  \NN
 \cmidrule(r){1-1}\cmidrule(rl){2-3}\cmidrule(rl){4-6}\cmidrule(l){7-7} 
 3a-01 & 2.40 & 1 & 2.39 & 1 & Partial & Pegasus                       \NN 
 3a-02 & 2.37 & 1 & 2.37 & 1 & Partial & Pisces
 \LL
}
\end{ctable}

Investigation of EIGs coordinates in the ``atlas of nearby large-scale structures'' \citep{1998Book..Fairall} shows that most EIGs reside close to walls and filaments rather than in centres of voids. This may explain why there is no EIG with $\dist{1} \geq 4.5\,\Mpch$ (as can be seen in Tables \ref{T:EIG12_Ngbr3Mpch} and \ref{T:EIG3_Ngbr3Mpch}).

\vspace{12pt}

A part of the Spring sky region, in which the $\alpha$.40 dataset covers a {3\,\Mpch} radius sphere around each point ($\mbox{8h} < {\RA} < \mbox{16h}$, $9\deg < {\Dec} < 11\deg$, $3500 < {\cz} < 7000\,\kms$) was statistically analysed. The number density of NED galaxies in this region is {0.065\,\NumCntMpch}, while the number density of ALFALFA galaxies in this region is {0.039\,\NumCntMpch}.
The average number of NED neighbours to a distance of {3\,\Mpch} around each NED galaxy in the above mentioned region was found to be $27.5 \pm 0.9$ (were the uncertainty is statistical and does not include the effect of uncertainties in {\cz}, which is expected to be minor). The average number of ALFALFA neighbours to a distance of {3\,\Mpch} around each NED galaxy in this region was found to be $14.3 \pm 0.5$. This is equivalent to a number density of $0.243 \pm 0.008\,\NumCntMpch$ NED neighbours, and $0.126 \pm 0.001\,\NumCntMpch$ ALFALFA neighbours, which means that the {3\,\Mpch} neighbourhood of randomly selected NED galaxies is 3 -- 4 times denser, on average, than the average density in the entire region. This result is expected, given the clustered nature of galaxy distribution in the Universe.

The ALFALFA neighbour counts (number of ALFALFA neighbours to a distance of {3\,\Mpch} listed in Table \ref{T:EIG12_Ngbr3Mpch}) of Spring galaxies that passed the criterion using NED data (subsamples EIG-1s and EIG-2s) and had full ALFALFA coverage were also statistically analysed. Their measured distribution fits well a Poisson distribution with an expected value of $0.7^{+0.4}_{-0.3}$ ALFALFA neighbours per EIG. Therefore, the average number density of ALFALFA neighbours within {3\,\Mpch} from these EIGs is \mbox{$0.006^{+0.004}_{-0.003}\,\NumCntMpch$}. This means that, on average, the number density around EIGs (1s and 2s) of ALFALFA galaxies is only $5\%^{+3\%}_{-2\%}$ of the number density around random NED galaxies.

\subsection{Comparison to other isolated galaxy samples}
\label{s:Envrnmnt_CompOther}

The ``Catalogue of Isolated Galaxies'' (CIG; \citealt{1973SoSAO...8....3K, 1986BICDS..30..125K}) was used as the basis of the AMIGA international project \citep{2007A&A...472..121V, 2013MNRAS.434..325F}. It defines a galaxy as isolated if it has no neighbours with angular diameter in the range $\tfrac{1}{4} a$ to $4 a$ up to a projected angular distance of $20a$, where $a$ is the angular diameter of the tested galaxy. 
\cite{2013MNRAS.433.1479H} estimated that a sample based on the AMIGA \citep{2007A&A...472..121V} 2D criterion will include a fraction of $\sim$18\% false positives due to projection effects.
For an angular diameter $a = 20\,\kpc$, for example, the \cite{1973SoSAO...8....3K} criterion corresponds to having no neighbours with angular diameters in the range 5 to {80\,\kpc} up to a distance of 0.4\,\Mpc.
Compared to this, the {4.26\,\,Mpc} distance criterion used in this work tests for a significantly higher level of isolation.

The same CIG 2D criterion was used for two other catalogues of IGs. 
The ``Two Micron Isolated Galaxy'' catalogue (2MIG) was created by \cite{2010AstBu..65....1K} from the ``Two Micron All-Sky Survey'' (2MASS; \citealt{2006AJ....131.1163S}) data using the selection criterion from CIG.
The ``Local Orphan Galaxies'' catalogue (LOG; \citealt{2011AstBu..66....1K}) was produced by combining a 3D redshift-space based criterion with the CIG 2D criterion. The LOG sample includes 520 IGs selected from a region defined by galactic latitudes $\left|b \right| >15\deg$ and with radial velocities smaller than {3500\,\kms} relative to the centroid of the Local Group. Their 3D criterion confirmed that the LOG sample galaxies are not part of gravitationally-bound groups that would survive the Hubble expansion. It assumed that 2MASS K-band luminosities are proportional to the total mass of galaxies. Their K-band luminosity-to-mass relation was tuned so that 10\% of the galaxies would pass the criterion (i.e.~would not be identified as part of a group).

The Void Galaxy Survey (VGS; \citealt{2012AJ....144...16K}) applied a redshift space criterion, very different from that used in this work. \cite{2012AJ....144...16K} used SDSS DR7 data to reconstruct the density field from the spatial galaxy distribution (in the redshift range $900 < \cz < 9000\,\kms$). Void regions were then identified in this density field using a ``watershed finder algorithm'' that does not assume a particular void size or shape. Sixty VGS sample galaxies were then selected to be as close as possible to the centres of these voids.

\vspace{12pt}

We tested the observed neighbourhoods of all galaxies of these four catalogues that are within the EIG search regions (defined in Table \ref{T:Sample-Regions}). The process and datasets used were identical to those used for the EIG selection. For each of the four catalogues, the probabilities of galaxies qualifying for each of the EIG subsamples (EIG-1, EIG-2 and EIG-3) are listed in Table \ref{T:Env_OtherFracEIG}.

\begin{ctable} 
[
  cap     = {Other IG catalogues - probability of qualification as EIGs},
  caption = {Other IG catalogues - probability of qualification as EIGs},
  star,
  label   = {T:Env_OtherFracEIG}
]
{@{}c@{\qquad}c@{\ }c@{\qquad}c@{\ }c@{\qquad}c@{\ }c@{\qquad}c@{\ }c@{}}
{
  \tnote[a]
  { 
    In square brackets are the numbers of tested IGs of each catalogue that qualify for each EIG subsample (or, under ``None'', that do not qualify for any subsample).
  }
}
{
 \FL
         & \multicolumn{8}{c}{Fraction {\footnotesize[number]}\tmark[a] qualifying as} \NN
 Catalogue & \multicolumn{2}{c}{EIG-1}                   & \multicolumn{2}{c}{EIG-2}
         & \multicolumn{2}{c}{EIG-3}                   & \multicolumn{2}{c}{None}   \ML
 AMIGA   & ($ 0^{+ 8}_{- 0} $)\% & {\footnotesize[0]}  & ($ 0^{+ 8}_{-0}$)\%  & {\footnotesize[0]}
         & ($ 7^{+12}_{- 5} $)\% & {\footnotesize[3]}  & ($93^{+ 5}_{-12}$)\% & {\footnotesize[40]} \NN
 2MIG    & ($ 0^{+ 7}_{- 0} $)\% & {\footnotesize[0]}  & ($ 0^{+ 7}_{-0}$)\%  & {\footnotesize[0]}
         & ($ 4^{+10}_{- 3} $)\% & {\footnotesize[2]}  & ($96^{+ 3}_{-10}$)\% & {\footnotesize[47]} \NN
 LOG     & ($ 0^{+26}_{- 0} $)\% & {\footnotesize[0]}  & ($ 0^{+26}_{-0}$)\%  & {\footnotesize[0]}
         & ($ 9^{+29}_{- 7} $)\% & {\footnotesize[1]}  & ($91^{+ 7}_{-29}$)\% & {\footnotesize[10]} \NN
 VGS     & ($33^{+37}_{-24} $)\% & {\footnotesize[2]}  & ($ 0^{+39}_{-0}$)\%  & {\footnotesize[0]}
         & ($50^{+31}_{-31} $)\% & {\footnotesize[3]}  & ($17^{+40}_{-14}$)\% & {\footnotesize[1]} \LL
}
\end{ctable}

As evident from the table, only a small fraction of the AMIGA, 2MIG and LOG catalogues may qualify as \mbox{EIG-3} galaxies (galaxies for which the distance to the closest neighbour in NED's data is {2 -- 3\,\Mpch}). None of the AMIGA, 2MIG and LOG galaxies \markChange{(within the regions defined in Table \ref{T:Sample-Regions})} fitted the EIG-1 or EIG-2 criterion, and none are part of the sample studied here (the EIG-3 subsample is not complete, i.e. does not include all galaxies that pass its criterion). 
\markChange
{
However, there is one 2MIG galaxy (outside the regions defined in Table \ref{T:Sample-Regions}), 2MIG 302, which is an EIG-1 galaxy (EIG 1a-04). It is not included in the statistics of Table \ref{T:Env_OtherFracEIG} since it lies outside the search region (as mentioned in section \ref{sec:Sample.TheIsoGals}). 
}

Only six galaxies from the VGS catalogue are within the search regions of the EIG sample. Three of these pass the criterion for the EIG-3 subsample, but are not part of it. One galaxy, VGS\_52, is an EIG-1 galaxy (EIG 1s-13). Another galaxy, VGS\_23, marginally qualifies as an EIG-1 galaxy and was not included in the sample studied here. The sixth VGS galaxy does not qualify for any of the EIG subsamples.

\begin{figure*}
\begin{centering}
\includegraphics[width=14cm,trim=0mm 0mm 0mm 0, clip]{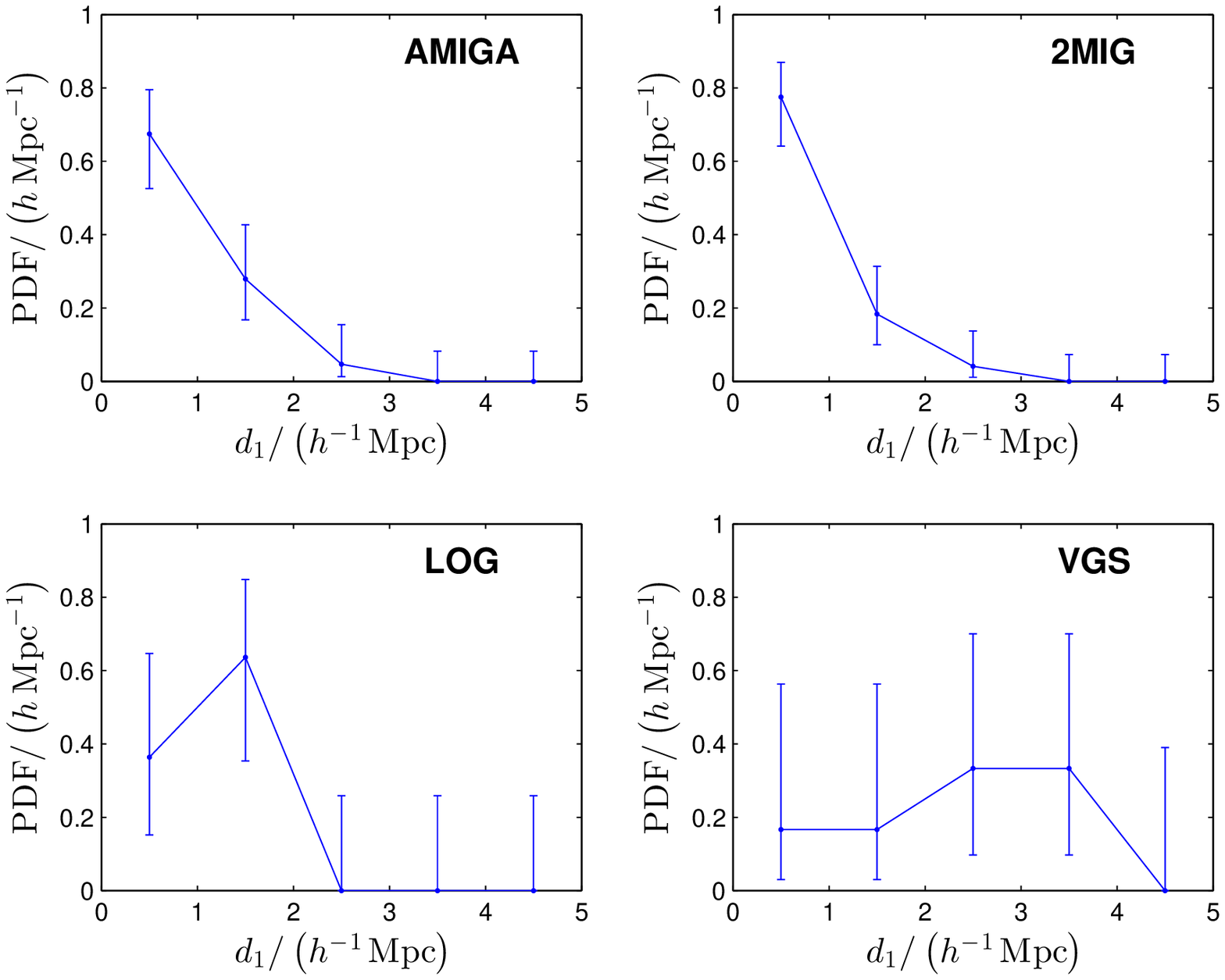}
\caption [PDF of \dist{1} for the AMIGA, 2MIG, LOG and VGS catalogues]
{
   PDF of the distance to the closest neighbour, \dist{1}, for the AMIGA, 2MIG, LOG and VGS isolated galaxy catalogues.\label{f:Env_Other_d1}
}
\end{centering}
\end{figure*}

\begin{figure*}
\begin{centering}
\includegraphics[width=14cm,trim=0mm 0mm 0mm 0, clip]{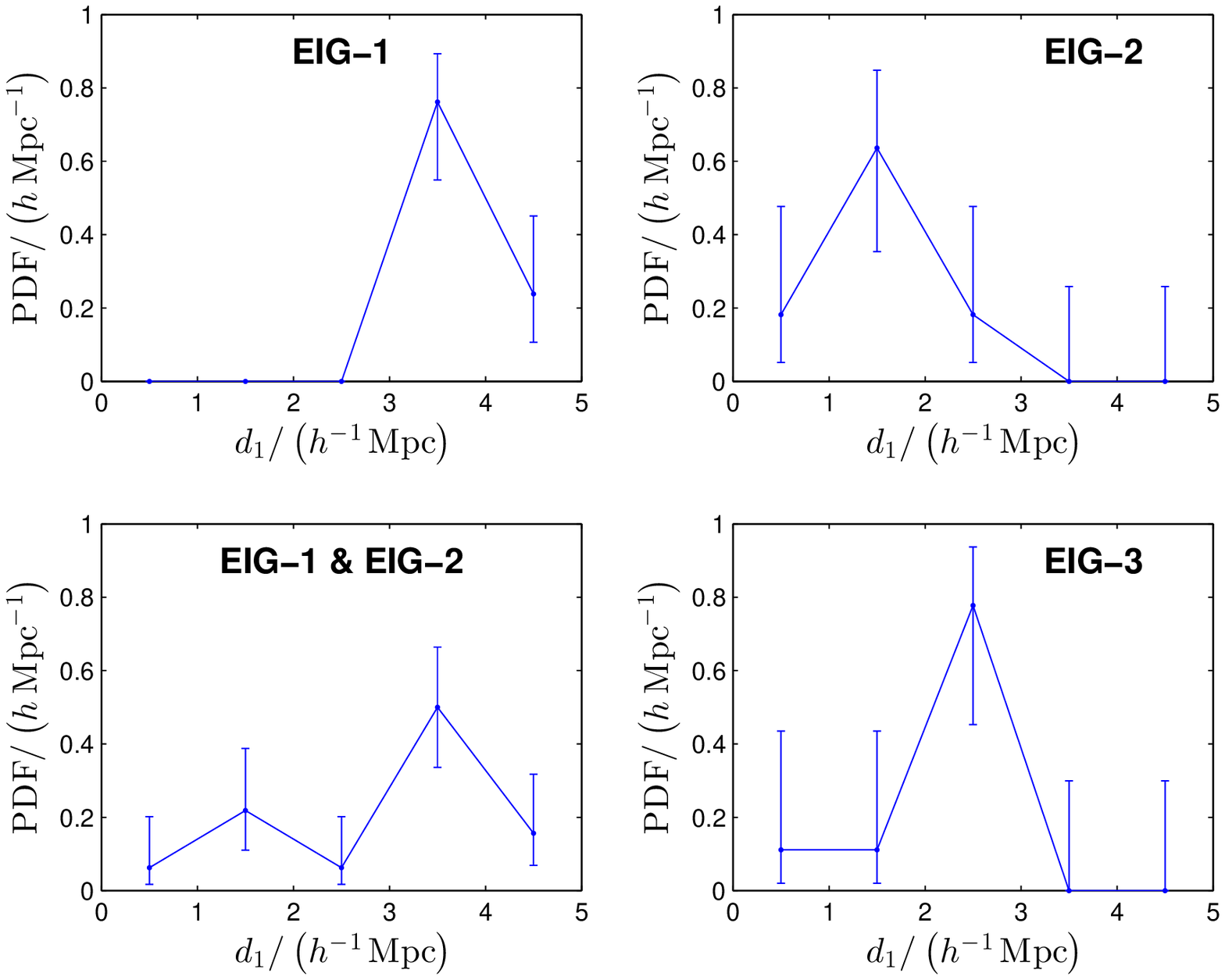}
\caption [PDF of \dist{1} for the EIGs]
{
   PDF of the distance to the closest neighbour, \dist{1}, for each subsample: ``EIG-1'', ``EIG-2'' and ``EIG-3'', and for the ``EIG-1 \& EIG-2'' subsamples together. \label{f:Env_EIG_d1}
}
\end{centering}
\end{figure*}

Distances to the closest neighbour listed in either the NED or $\alpha$.40 datasets, \dist{1}, were measured for all AMIGA, 2MIG, LOG and VGS galaxies in the EIG search regions. Based on these, the probability distribution function (PDF) of \dist{1} was calculated for each catalogue (Figure \ref{f:Env_Other_d1}). For comparison, Figure \ref{f:Env_EIG_d1} shows the PDF of \dist{1} for each of the EIG subsamples and for the EIG-1 and EIG-2 subsamples together (all galaxies that passed the isolation criterion using the NED dataset).

It is evident from these figures that the \dist{1} of AMIGA, 2MIG and LOG galaxies is typically significantly lower than the \dist{1} of EIG galaxies. The average \dist{1} of the tested galaxies was {0.83\,\Mpch}
 for AMIGA, {0.74\,\Mpch} for 2MIG and {1.19\,\Mpch} for LOG, \markChange{compared to {3.54\,\Mpch} for the EIG-1 subsample, {1.58\,\Mpch} for the EIG-2 subsample and {2.39\,\Mpch} for the EIG-3 subsample. The average \dist{1} of the EIG-1 and EIG-2 subsamples together is {2.86\,\Mpch}.}
The PDF for the VGS catalogue reaches higher \dist{1} values compared to the other three catalogues. The average \dist{1} measured for the six tested VGS galaxies is {2.39\,\Mpch}.

\vspace{24pt}

We conclude that the EIG sample studied here is indeed extreme in its measurable isolation. The use of HI redshifts from ALFALFA proved to be a key factor in identifying the most extremely isolated subsample (EIG-1).

%% file: Chapters/3_Environment.tex
\section{Properties Estimated using Cosmological Simulations}
\label{s:Envrnmnt_Simulations}

This section describes the analysis of cosmological simulations we performed to estimate properties of the EIG-1 and EIG-2 subsamples.
Two cosmological simulations were used for this analysis (described in section \ref{s:SimSimulations}) using which mock EIG samples were created (sections \ref{s:completenessFunc} through \ref{s:SimMockEIGs}). By comparing these with random mock samples, properties of the EIGs were statistically estimated.

In section \ref{s:SimHaloProp} properties of the dark matter (DM) haloes are analysed. These include the halo's mass and whether it is dominant in its immediate neighbourhood. In Section \ref{s:SimHaloHist} mass accretion histories (MAHs) are analysed. Next, the neighbourhoods of the EIGs are analysed in terms of galaxy number density (section \ref{s:SimNgbrhdGals}), halo number density (section \ref{s:SimNgbrhdHalos}) and DM mass density (section \ref{s:SimNgbrhdMassDens}). Finally, the tidal acceleration exerted on EIGs by their neighbouring DM haloes is analysed (section \ref{s:SimTide}).

\vspace{12pt}

As discussed in section \ref{sec:Sample.Criterion}, the neighbourhood measurement is limited by two factors: incompleteness of the redshift data (i.e., redshift data is not available for a significant fraction of the galaxies), and peculiar velocities that introduce an error in the distance measurement. Due to these, the actual neighbourhood of an individual EIG may differ significantly from what it seems to be from the data in Tables \ref{T:EIG12_Ngbr3Mpch} and \ref{T:EIG3_Ngbr3Mpch}, or from the number density functions, such as shown in Figure \ref{f:Sample-n_r_3Gals}.

However, as a sample, rather than individually, the probabilities of neighbourhood properties can be derived using cosmological simulations. These simulations describe mock universes with detailed information on DM haloes and galaxies that reside in them. By applying the same search process used previously to select the EIG sample on these mock universes, mock EIG samples were created for which the simulated properties were calculated. The distribution of these properties in the mock EIG samples serves as an estimate of the probability distribution functions (PDFs) of these properties in the real EIG sample.

The derivation process of the PDFs included the following steps:

\begin{itemize}
  \item Defining points of view and sky regions in the mock universes, simulating the observer and the sky region in which the mock EIGs are searched for.

  \item Estimating ``completeness'' functions of the NED data, which define for each given observable magnitude, the fraction of galaxies in the search region for which a redshift measurement was available in the NED dataset. The ``completeness'' functions of the Spring and Autumn regions were measured separately, since they are significantly different (the Spring region is fully covered by SDSS, while the Autumn region is not). The ``completeness'' functions were estimated separately for each of the simulations.

  \item Creating ``mock observable'' datasets, each including all coordinates and simulation IDs of galaxies, randomly selected using the ``completeness'' function. These ``mock observable'' datasets imitate the data that would have been available from NED, had the ``mock universes'' been the real Universe. For each simulation, point of view, and sky region, several such ``mock observable'' datasets were created.

  \item Creating ``mock EIG samples'' by applying the sample selection process (described in section \ref{s:SampleProcess}) on the ``mock observable'' datasets. These are divided to ``Spring mock EIG samples'', which simulate subsamples EIG-1s and EIG-2s (together), and ``Autumn mock EIG samples'', which simulate subsamples EIG-1a and EIG-2a (together).

  \item Creating ``mock random samples'' by randomly selecting a thousand galaxies from each ``mock observable'' dataset. These ``mock random samples'' are used as reference to the ``mock EIG samples'', when evaluating their properties' PDFs.

  \item ``Measuring'' simulated properties of ``mock EIG samples'' and ``mock random samples'' galaxies, and creating histograms that estimate the PDFs of these properties in the real Universe EIG sample, and in real Universe random galaxies.

\end{itemize}
\vspace{12pt}

Note that, at the time this analysis was performed no cosmological simulation claimed to estimate the HI content of galaxies with reasonable accuracy\footnote{The only simulation that estimates HI content to date is Illustris \citep{2014Natur.509..177V}. Its data became public only on April 2015 \citep{2015A&C....13...12N}, thus Illustris was not considered for the analysis presented here.}. Therefore, the ``completeness'' functions were defined for the luminous content only, and not for HI content (21cm fluxes). The estimated PDFs discussed below, therefore, relate more closely to the EIG-1 and EIG-2 subsamples together (all galaxies that passed the isolation criterion using the NED dataset), rather than to each subsample separately.

As already shown in section \ref{sec:Sample.TheIsoGals}, the use of ALFALFA significantly improves the quality of the sample. Therefore, the isolation properties of the EIG-1 subsamples are expected to exhibit significantly more isolated-like PDFs compared to the PDFs estimated here (for EIG-1 and EIG-2 together).

\vspace{12pt} 


\subsection{Simulations}
\label{s:SimSimulations}

The following two cosmological simulations were used independently for the EIGs history and neighbourhood analysis.

\subsubsection{Millennium II-SW7 (Mill2)}

The Millennium II-SW7 simulation (Mill2; \citealt{2013MNRAS.428.1351G}) made publically available by the Virgo Consortium \citep{2006astro.ph..8019L} is an updated version of the Millennium-II simulation \citep{2009MNRAS.398.1150B} in which the structure growth in a $\Lambda$ cold dark matter ($\Lambda$CDM) universe was scaled to parameters consistent with WMAP7 \cite{2011ApJS..192...17B}. The properties of galaxies were simulated using the semi-analytical model (SAM) described in \cite{2013MNRAS.428.1351G}.

Mill2 simulates a cube with edge length of $104.311\,\Mpch$, and uses $2160^3$ particles of mass $\Mass_{p} = 8.5024 \cdot 10^{6}\,\Msunh$ each.
It uses the following cosmological parameters: $\h = 0.704$, $\Omega_{\Lambda} = 0.728$ (density parameter for dark energy), $\Omega_{m} = 0.272$ (density parameter for matter), $\Omega_{b} = 0.045$ (density parameter for baryonic matter), $n_{s} = 0.961$ (normalization of the power spectrum), and $\sigma_{8} = 0.807$ (amplitude of mass density fluctuation in {8\,\Mpch} sphere at $\z = 0$).

Two types of halo classifications are defined in Mill2:
\begin{itemize}
  \item Friends of Friends (FOF) groups
 - defined with $b = 0.2$ \citep{2009MNRAS.398.1150B}\footnote{In the FOF method \citep{1985ApJ...292..371D} all particle pairs separated by less than a fraction, $b$, of the mean interparticle separation (linking length) are found. Each distinct subset of connected particles is then defined as an FOF group.}.

  \item (Sub)Haloes - The decomposition of the FOF groups into gravitationally-bound haloes.
\end{itemize}

\vspace{12pt}

For each subhalo, a merger tree can be extracted from the simulation, which includes data on all its progenitors since the beginning of the simulated time (lookback time of {13.75\,\Gyr}). The merger tree and physical properties of each progenitor subhalo are the inputs of the SAM. 
The halo dataset used for the analysis described in this work was limited by halo mass $\Mhalo \geq 10^{9}\,\Msunh$. \rem{Haloes of lower masses were not included in the calculations. This might somewhat affect the values obtained for the neighbourhood mass density and for the tidal acceleration (sections \ref{s:SimNgbrhdMassDens} and \ref{s:SimTide}) but is expected to do so similarly for the EIGs and for the random mock samples.}

As described above at the beginning of section \ref{s:Envrnmnt_Simulations}, the analysis required choosing a point of view (simulating our Galaxy) and its tested sky region. Simulated EIGs were then searched for in this sky region, using data in the redshift range {1600 -- 7400\,\kms} (same as used for the search in the NED dataset). Mill2's simulated box is too small to allow a $4\pi\,\sr$ coverage to this range. To simplify the search algorithm (avoiding the use of the simulation's periodic boundary conditions) the point of view was chosen to be close to the simulation's point of origion (one of the corners of the simulated cube) at $\vec{r} = \left(20.0\,,~20.0\,,~20.0\right)\,\Mpch$, and the tested sky region was chosen to be the (+,+,+) octant. The {20.0\,\Mpch} distance in each axis from the cube's corner was chosen to allow a simplified search around the edges of the search region.


\subsubsection{Box160}

The Box160 simulation is a constrained simulation of the local Universe, based on the $\Lambda$CDM third-year WMAP (WMAP3, \citealt{2007ApJS..170..377S}) which simulates a cube with $160\,\Mpch$ edges \citep{2008arXiv0803.4343G, 2009MNRAS.396.1815F}. It is part of the Constrained Local UniversE Simulations project (CLUES, \citealt{2010arXiv1005.2687G}).  Its DM distribution emulates large structures in the local Universe (Virgo, Coma, Local Supercluster, etc.).
Box160 uses $1024^{3}$ DM particles each of mass $\Mass_{p} = 2.54 \cdot 10^{8} \Msunh$, and the following cosmological parameters: $\h = 0.73$, $\Omega_{\Lambda} = 0.76$, $\Omega_{m} = 0.24$ , $n_{s} = 0.96$, and $\sigma_{8} = 0.76$.

Unlike Mill2, in Box160 the FOF haloes are not divided into gravitationally-bound subhaloes. Instead, the FOF haloes are directly populated with simulated galaxies (a single FOF halo may contain more than one galaxy). The algorithm applied for this is a conditional luminosity function (CLF) algorithm similar to that of \cite{2007MNRAS.376..841V} but without distinction between central and satellite galaxies (which may somewhat alter the probability that close central-satellite pairs will be detected as non-isolated).
The faintest simulated galaxy luminosity is $3.305 \cdot 10^7 \Lsunh$, corresponding to $\AbsMg = -14.44$.
The Box160 dataset used here contains haloes of mass $1.814 \cdot 10^{10}\,\Msunh$ and above.

For the analysis of Box160 two points of view were used from which the neighbourhood resembles that of our Galaxy (S. Gottloeber \& Y. Hoffman, private communication). These points, around which the entire $4\pi\,\sr$ sky was tested, are defined by their location, $\vec{r}$, and by their peculiar velocity (relative to the comoving coordinates), $\vec{v_{p}}$, as follows:
\begin{description}
  \item[LG1] :   $\vec{r}    = \left(79.3241 \,,~ 56.0739 \,,~ 84.7330 \right) \,\Mpch\ ,$
   \item[\qquad \;\,] $\vec{v_{p}} = \left(28.7    \,,~ 427.8   \,,\, -340.1  \right) \,\kms$
  \item[LG2] :   $\vec{r}    = \left(74.9237 \,,~ 63.8382 \,,~ 80.4162 \right) \,\Mpch\ ,$
   \item[\qquad \;\,] $\vec{v_{p}} = \left(-172.4  \,,~ 511.2   \,,\, -349.5  \right) \,\kms$
\end{description}


\vspace{18pt}

Compared to the Mill2 simulation, Box160 is inferior in mass resolution and in the fact that it uses older cosmological parameters. However, since it simulates the local Universe, it enables analyzing the isolation criterion from a point of view resembling ours in the real Universe. This, along with the differences in halo definition and method of populating the haloes with galaxies, serves as a tool for estimating the sensitivity of the results to these important simulation details.

%
%
%
%

%
%
%
%
%
%

\subsection{The completeness functions}
\label{s:completenessFunc}

In order to estimate the completeness function (described at the beginning of section \ref{s:Envrnmnt_Simulations}) the number density of galaxies per magnitude interval, $n_{gal} / \Delta m$, was derived both for the simulations and for the NED datasets. For the simulations, $n_{gal} / \Delta m$ was calculated for the same redshift range as that of the NED datasets, i.e.~{1600 -- 7400\,\kms}. 

\begin{figure*}
\begin{centering}
\includegraphics[width=14cm,trim=0mm 0mm 0mm 0, clip]{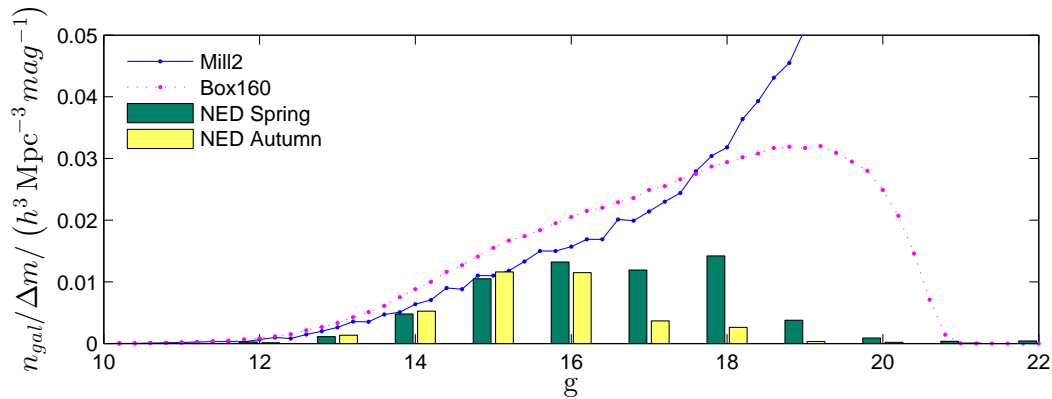}
\caption [Galaxy number density per magnitude, $n_{gal} / \Delta m$, of the simulations and NED data.]
{ 
  The number density of galaxies per magnitude interval, $n_{gal} / \Delta m$, for the Mill2 simulation, Box160 simulation (average between LG1 and LG2), NED's Spring region, and NED's Autumn region (redshift range: $1600 \leq \cz \leq 7400\,\kms$). \label{f:sim_n_gal_mag}
}
\end{centering}
\end{figure*}

Figure \ref{f:sim_n_gal_mag} shows the galaxy number density per magnitude interval, $n_{gal} / \Delta m$, for the Mill2 simulation, Box160 simulation, NED's spring region, and NED's Autumn region. For Box160 $n_{gal} / \Delta m$ was measured from both points of view (LG1 and LG2). The results for LG1 and LG2 were very similar (less than $9\cdot10^{-4}\,\NumCntMpch\,mag^{-1}$ apart at any point). The curve displayed in the figure for Box160 is the average between them.

As Figure \ref{f:sim_n_gal_mag} shows, $n_{gal} / \Delta m$ of Mill2 and Box160 are somewhat different for $\SDSSg < 18$. The Box160 curve peaks at $\SDSSg \cong 19$, while the Mill2 curve continues to rise and peaks only at $\SDSSg = 28.7$ with a value of $\sim 0.7\,\NumCntMpch\,mag^{-1}$. These differences are believed to be mainly due to the different method by which these simulations populate haloes with galaxies. The SAM used by Mill2 can create extremely low-luminosity galaxies, effectively extrapolating far beyond the accurately measured range of the galaxy luminosity function (LF). Box160 on the other hand, populates haloes by imitating the measured LF, and therefore cannot extrapolate beyond the LF established range.

It is also evident from Figure \ref{f:sim_n_gal_mag} that $n_{gal} / \Delta m$ of the NED Spring region differs significantly from that of the NED Autumn region. The fact that $n_{gal} / \Delta m$ of the Autumn region is cut off at brighter magnitudes compared to the Spring region hints that this is due to completeness differences, where the Spring region is complete to fainter magnitudes. This is probably because the Spring region is fully covered spectroscopically by SDSS, while the Autumn region is not. However, the difference in $n_{gal} / \Delta m$ may also be attributed to a real difference in the large-scale structure of these two regions; the density of faint galaxies in the Autumn region may really be lower in comparison to the Spring region.
The analysis of the simulations assumes an average composition of the large-scale structures. Any deviation from an average composition (in the Spring or Autumn regions) affects the accuracy of the completeness function estimation, which propagates to the accuracy of the calculated PDFs.
Evidence for such a deviation is discussed in section \ref{s:SimMockEIGs}.

The completeness functions, defining the fraction of galaxies for which a redshift measurement was available in the NED dataset, were fit to the following simple two parameter model:
\begin{equation}
\mbox{completeness function} \left( \SDSSg \right) =  
  \begin{cases}
        \mbox{frac}_{obs} &  \SDSSg \leq \SDSSg_{max} \\
        0           &  \SDSSg >    \SDSSg_{max}
  \end{cases} \end{equation}
where: 
\begin{description}
  \item[$\SDSSg_{max}$] is the cutoff {\SDSSg} magnitude above which no galaxy is observed 
  \item[\normalfont $\mbox{frac}_{obs}$] is the fraction of galaxies observed below the cutoff magnitude.
  \item[]
\end{description}

This model emulates typical spectroscopic surveys in which the redshifts of galaxies dimmer than a limiting magnitude ($\SDSSg_{max}$) are not measured at all, while not all of the brighter galaxies are measured. This is obviously not an accurate model for datasets such as NED, which include a combination of many spectroscopic surveys.

One alternative to this simplified completeness model is to apply a completeness correction for each magnitude bin separately. This would trace more tightly the $n_{gal} / \Delta m$ of NED's Spring and Autumn regions compared to the simplified model used here. However, this might cause inaccuracies when overdensities or underdensities in certain magnitude bins (due to, for example, variation from the average composition of the large-scale structures) will be falsely translated to overestimates or underestimates in the completeness function.

The variables $\mbox{frac}_{obs}$ and $\SDSSg_{max}$ were calculated to provide simultaneous fits to the following two parameters:
\begin{itemize}
 \item The galaxy number density ($n_{gal} / \Delta m$ integrated over all magnitudes) and
 \item $n_{gal} / \Delta m$ integrated over the decreasing slope ($\SDSSg \geq 17.5$ for the Spring region, and $\SDSSg \geq 15.5$ for the Autumn).
\end{itemize}

This ensures that the overall number density, as well as the number density of the dimmer galaxies, are well-simulated in the mock observable datasets.

The best-fitted parameters of the completeness function are listed in Table \ref{T:SimCompParams}. As can be seen, $\SDSSg_{max}$ is significantly larger for the Spring region, while $\mbox{frac}_{obs}$ is somewhat larger for the Autumn region. There are large differences between the parameters of the Mill2 simulation and those of the Box160 simulation. These are the result of the differences in the $n_{gal} / \Delta m$ functions, discussed above.

\begin{ctable}
[
  cap     = {The completeness functions},
  caption = {The completeness functions' fitted parameters},
  label   = {T:SimCompParams}
]
{@{}ccp{0.3cm}cc@{}}
{
}
{
 \FL
 Region                & Simulation && $\mbox{frac}_{obs}$ & $\SDSSg_{max}$
 \ML
 \multirow{2}*{Spring} & Mill2      && 71.4\%              & 18.4 \NN
                       & Box160     && 57.4\%              & 18.7 
 \ML
 \multirow{2}*{Autumn} & Mill2      && 86.3\%              & 16.8 \NN
                       & Box160     && 63.5\%              & 16.9
 \LL
}
\end{ctable}

\subsection{The mock observable datasets}
\label{s:NgbrMockObs}

The ``Mock Observable'' datasets were created using the procedure described at the beginning of section \ref{s:Envrnmnt_Simulations} and using the completeness functions described in section \ref{s:completenessFunc}. For the Mill2 simulations five mock observable datasets were created for the Spring region, and five for the Autumn region. For the Box160 simulation three mock observable datasets were created for each point of view (LG1 and LG2) and region, for a total of six mock observable datasets for the Spring region, and six for the Autumn region.

\begin{figure*}
\begin{centering}
\includegraphics[width=13cm,trim=0mm 0mm 0mm 0, clip]{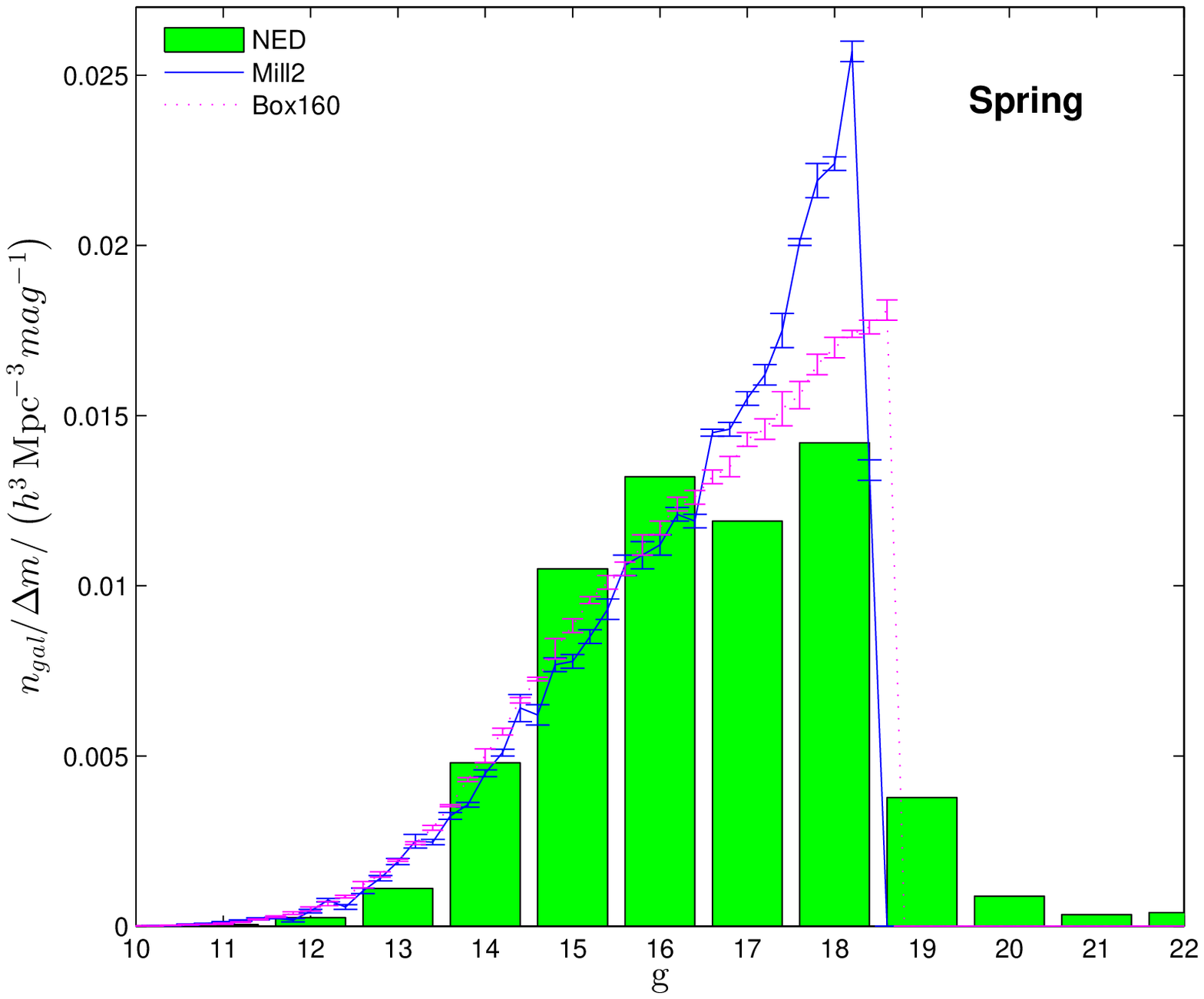}
\includegraphics[width=13cm,trim=0mm 0mm 0mm 0, clip]{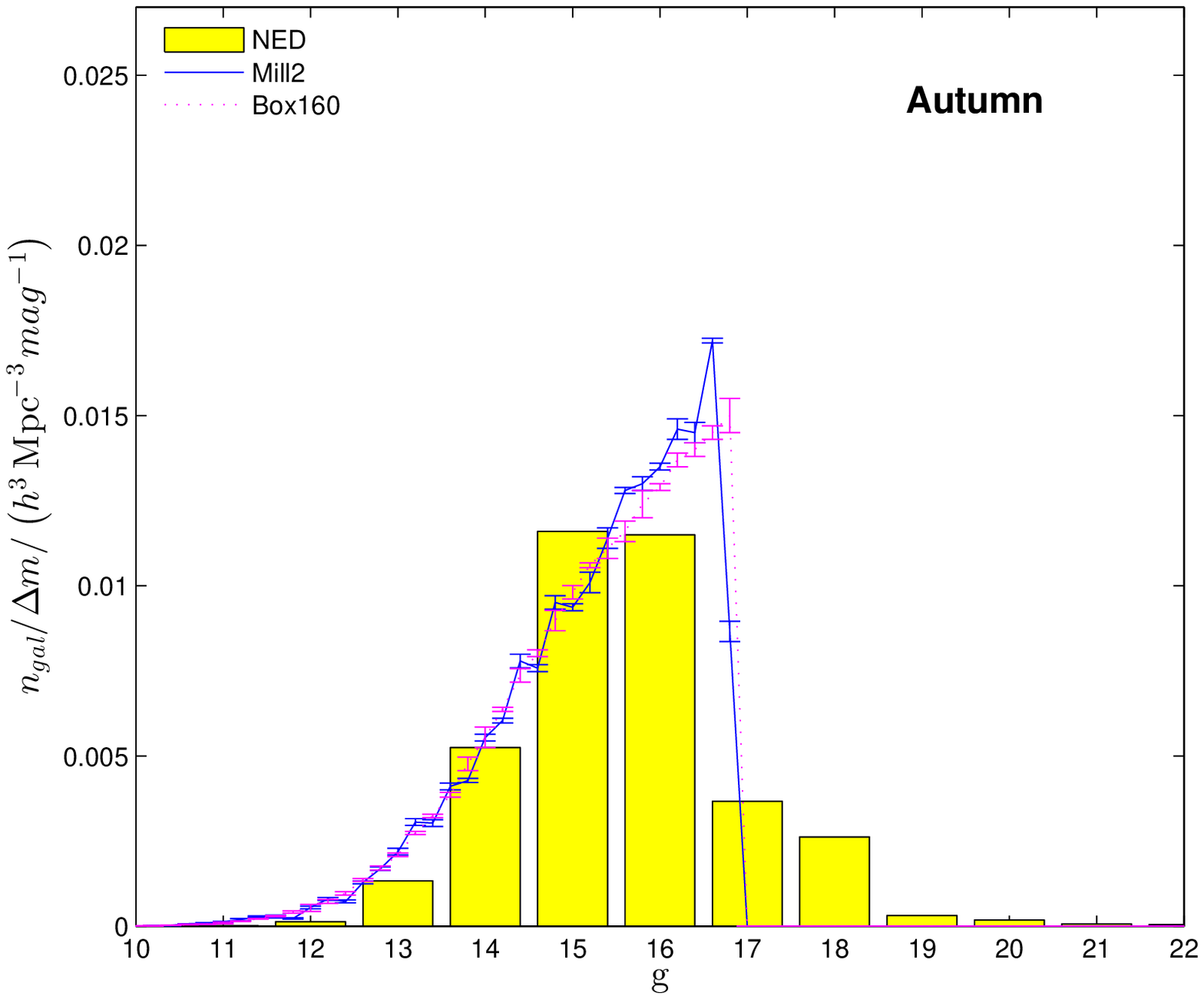}
\caption [A comparison of $n_{gal} / \Delta m$ of the Mill2 mock observable datasets, Box160 mock observable datasets, and NED.]
{ 
  A comparison of $n_{gal} / \Delta m$ of the Mill2 mock observable datasets, Box160 mock observable datasets, and NED (in the redshift range: $1600 \leq \cz \leq 7400\,\kms$).\rem{Left panel for the Spring sky region. Right panel for the Autumn region.}\label{f:Sim_Completement_test}
}
\end{centering}
\end{figure*}

For each mock observable dataset, the $n_{gal} / \Delta m$ function was calculated. The functions were averaged for each of the four simulation and sky region pairs. These average functions are shown in Figure \ref{f:Sim_Completement_test}, along with the NED $n_{gal} / \Delta m$ to which they were fitted. The vertical error bars in the figure show the standard deviation of the averaged functions.

As Figure \ref{f:Sim_Completement_test} shows, the mock observable datasets trace approximately the real NED $n_{gal} / \Delta m$ function. In the range $\SDSSg < 16$ the fit is reasonably well. In the ranges $16 < \SDSSg < 18$ for the Spring region, and $16 < \SDSSg < 17$ for the Autumn region, there is an excess of simulated galaxies. Due to the method by which the completeness function was fitted, this excess averages with the deficiency above the cutoff magnitude, $\SDSSg_{max}$, in which by definition there are no simulated galaxies so that the overall number densities of galaxies in the simulations are the same as in NED.

\subsection{The mock EIG samples}
\label{s:SimMockEIGs}

``Mock EIG samples'' were created by applying the search criterion described in section \ref{sec:Sample.Criterion} to each mock observable dataset.
The fraction of galaxies with known redshifts that passes the isolation criterion, $\mbox{frac}_{EIG}$, and the number density of the EIGs, $n_{EIG}$, were calculated for each mock EIG sample and for the real NED data. Table \ref{T:SimEIGfrac_n} presents the results. For each simulation and sky region, the average values are shown with standard-deviation-based ($1\sigma$) uncertainties. The table also lists the number of mock random galaxies, $N_{random}$, and EIGs, $N_{EIG}$, that were analysed. 


\begin{ctable} 
[
  caption = { EIG abundance },
  star,
  label   = {T:SimEIGfrac_n}
]
{@{}cccc@{\quad}cc@{}}
{
}
{
 \FL
   Region                & Dataset & $N_{random}$ & $N_{EIG}$ & $\mbox{frac}_{EIG}$ 
                                         & $n_{EIG} \left[ h^{3}\,\mbox{Mpc}^{-3} \right]$
 \ML
   \multirow{3}*{Spring} & NED     &   ---        & \markChange{21} & $\left(0.62 \pm 0.14 \right) \% $
                                                  & $\left( 3.9  \pm 0.8 \right) \cdot 10^{-4}$  \NN
                         & Mill2   &  5000        &   522     & $\left(0.97 \pm 0.07 \right) \% $
                                                  & $\left( 5.0  \pm 0.4 \right) \cdot 10^{-4}$  \NN
                         & Box160  &  6000        &  4952     & $\left(0.90 \pm 0.03 \right) \% $
                                                  & $\left( 4.91 \pm 0.07\right) \cdot 10^{-4}$ 
 \ML
   \multirow{3}*{Autumn} & NED     &   ---        &    10     & $\left(2.6  \pm 0.8  \right) \% $
                                                  & $\left(11    \pm 3   \right) \cdot 10^{-4}$  \NN
                         & Mill2   &  4000        &   625     & $\left(1.96 \pm 0.05 \right) \% $
                                                  & $\left( 6.0  \pm 0.2 \right) \cdot 10^{-4}$  \NN
                         & Box160  &  6000        &  6449     & $\left(1.99 \pm 0.09 \right) \% $
                                                  & $\left( 6.40 \pm 0.15\right) \cdot 10^{-4}$
 \LL
}
\end{ctable}

As Table \ref{T:SimEIGfrac_n} shows, the simulations reproduce the real EIG abundance \markChange{to within $\sim$2\,$\sigma$} in both measured parameters.
The abundance of Autumn EIGs is significantly larger than the abundance of Spring EIGs. The Autumn-to-Spring abundance ratio is larger in the real NED data ($4 \pm 2$ in $\mbox{frac}_{EIG}$, and $3 \pm 1$ in $n_{EIG}$) compared to the simulations (a factor of $2.17 \pm 0.08$ in $\mbox{frac}_{EIG}$, and $1.26 \pm 0.03$ in $n_{EIG}$). This may be attributed to a real difference between the large-scale structure of the Autumn and Spring regions, which is not accounted for in the simulations.

\subsection{Halo properties}
\label{s:SimHaloProp}

Extremely isolated galaxies are expected to reside in (sub)haloes that are the dominant ones in their FOF group, i.e. haloes that do not have neighbours of comparable mass within the FOF group.
The Mill2 simulation was analysed to find out which of the galaxies reside in such dominant subhaloes, where a dominant subhalo is defined here as one lacking subhalo neighbours in its FOF halo with mass $\geq 25\%$ of its own. The fraction of galaxies residing in dominant subhaloes, $\mbox{frac}_{dominant}$, was calculated for the mock EIG samples and for the mock random samples in both sky regions. The results are presented in Table \ref{T:SimMill2DomFrac}.

\begin{ctable} 
[
  caption = {Fraction dominant in FOF halo},
  mincapwidth = 90mm,
  label   = {T:SimMill2DomFrac}
]
{@{}c@{\quad}c@{\quad}c@{}}
{
}
{
 \FL
   Region                & Sample & $\mbox{frac}_{dominant}$
 \ML
   \multirow{2}*{Spring} & EIG     & $0.94 \pm 0.02$  \NN
                         & Random  & $0.54 \pm 0.01$
 \ML
   \multirow{2}*{Autumn} & EIG     & $0.91 \pm 0.01$  \NN
                         & Random  & $0.58 \pm 0.02$
 \LL
}
\end{ctable}

Table \ref{T:SimMill2DomFrac} shows that the fraction of EIGs that are dominant in their FOF haloes is significantly larger than the overall fraction of dominant galaxies (``measured'' in the random samples). This dominant fraction is slightly larger for Spring EIGs than for Autumn EIGs, probably due to the higher completeness of the Spring data. The Autumn overall dominant fraction is slightly larger than the Spring one, probably due to the fact that the Autumn mock observable datasets contain fewer dim galaxies, which are more abundant in satellite subhaloes.

\vspace{12pt}

\begin{figure*}
\begin{centering}
\includegraphics[width=14cm,trim=0mm 0mm 0mm 0, clip]{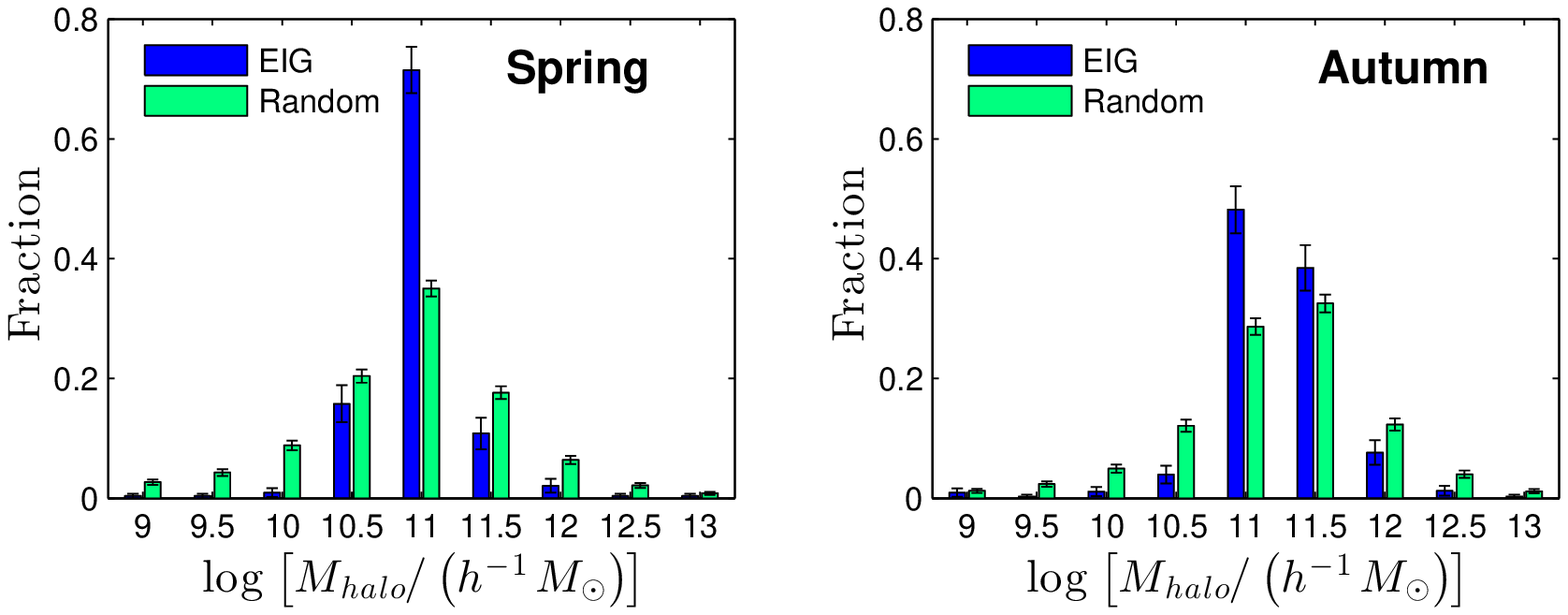}
\caption [Mill2 - subhalo mass distribution]
{
  Distribution of subhalo mass, {\Mhalo}, in Mill2 mock datasets (EIG and random).\rem{Left panel for the Spring sky region. Right panel for the Autumn region. }\label{f:Sim_Mill2HaloMass}
}
\end{centering}
\end{figure*}

The PDFs of the galaxies' subhalo mass, {\Mhalo}, were ``measured'' in Mill2 and are shown in Figure \ref{f:Sim_Mill2HaloMass}. The figure compares the {\Mhalo} PDF of EIGs with that of randomly selected galaxies, for both the Spring (left panel) and Autumn (right panel) regions.
Figure \ref{f:Sim_Mill2HaloMass} shows that the {\Mhalo} of the EIGs is less scattered than that of the random population. This is more pronounced in the Spring region, where the standard deviation of $\log \left[ \Mhalo / \left( \Msunh \right) \right]$ is $\sim$0.3 for the EIGs, compared to $\sim$0.8 for the random sample (for the Autumn region it is $\sim$0.4 for the EIGs, compared to $\sim$0.7 for the random sample).

\begin{figure*}
\begin{centering}
\includegraphics[width=14cm,trim=0mm 0mm 0mm 0, clip]{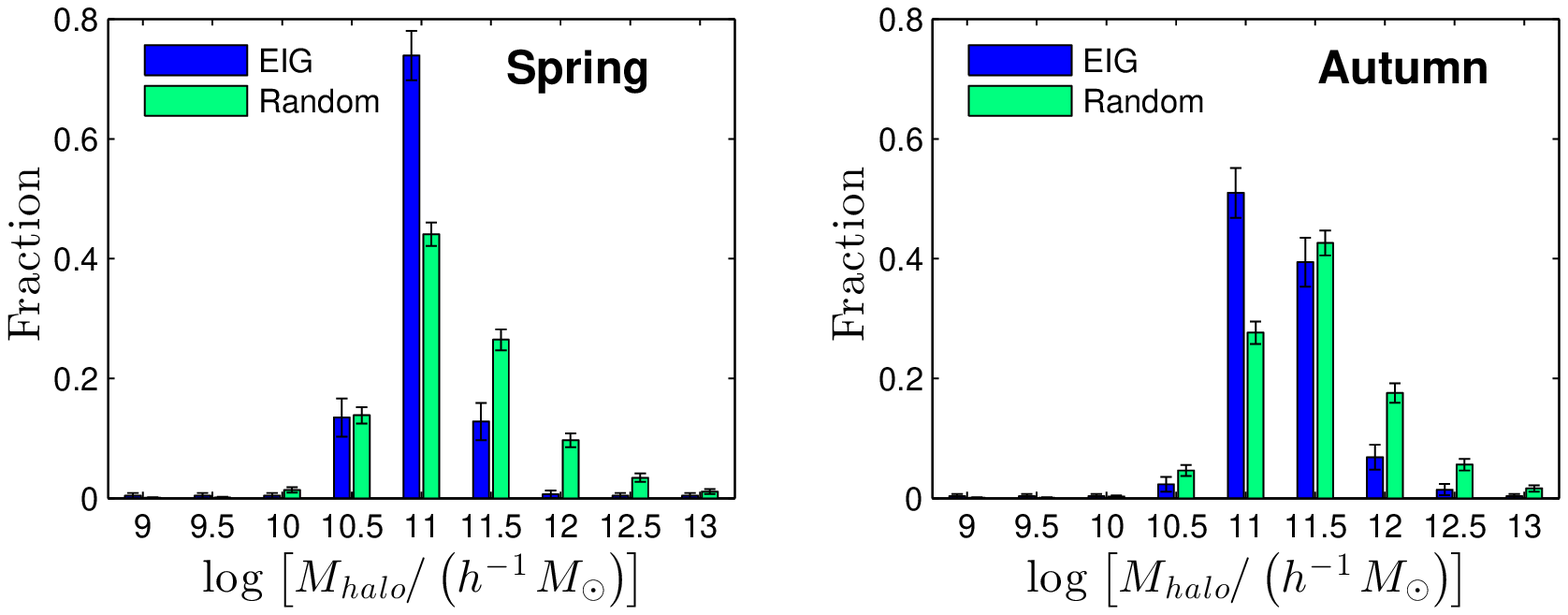}
\caption [Mill2 - subhalo mass distribution of haloes that did not experience a ``major event'']
{
  Distribution of subhalo mass, {\Mhalo}, in Mill2 mock datasets (EIG and random) of haloes that did not experience a ``major event'' in their past. \rem{ Left panel for the Spring sky region. Right panel for the Autumn region. }\label{f:Sim_Mill2HaloMassWO_MjrEvnt}
}
\end{centering}
\end{figure*}

Interestingly, there's no significant difference in the average $\log \left[ \Mhalo / \left( \Msunh \right) \right]$ between the EIG and the random samples. In the Spring region it is 11.0 (EIGs) vs.~10.9 (random). In the Autumn region it is 11.3 (EIGs) vs.~11.2 (random).
The difference between the two regions can be explained by the Spring region completeness function that reaches fainter magnitudes. The fainter galaxies included in the Spring dataset typically reside in lower mass haloes.

The Mill2 simulation, therefore, predicts that isolated galaxies tend to reside in haloes of average mass, rather than in low or high-mass haloes. The fraction of EIGs in haloes of $\Mhalo > 10^{12}\,\Msunh$ or $\Mhalo < 10^{10}\,\Msunh$ is significantly smaller than this fraction in a random population. The low abundance of EIGs with $\Mhalo > 10^{12}\,\Msunh$ may be attributed to the low mass density in the isolated regions where EIGs reside, which possibly does not allow them to accrete so much mass.

One may speculate that for $\Mhalo < 10^{10}\,\Msunh$ this could be the result of a selection effect, if EIGs in low mass haloes had significantly lower luminosities compared to random galaxies of the same halo mass. In such case, the EIGs of low halo mass would not have been detectable due to their extremely low luminosities. However, this does not seem to be the case. A comparison between the absolute magnitudes of the EIGs and random galaxies of the same mass bins shows that in the central bins ($10^{10.5}\,\Msunh$ to $10^{12}\,\Msunh$) the average difference is much smaller than its standard deviation, and would not explain this phenomenon. For the lower mass bins, the few EIGs found in Mill2 even show a tendency to be brighter than the random galaxies.

The explanation for the low abundance of EIGs with $\Mhalo < 10^{10}\,\Msunh$ seems to be related to the low rate of ``major events'' they undergo (major mergers or significant mass loss events), as defined and analysed in section \ref{s:SimHaloHist}.
This is evident from a comparison between Figure \ref{f:Sim_Mill2HaloMass} and Figure \ref{f:Sim_Mill2HaloMassWO_MjrEvnt}, which shows the {\Mhalo} PDF of subhaloes that did not experience a ``major event'' in their past. In the range $\Mhalo \leq 10^{10}\,\Msunh$, the PDF of random galaxies that did not experience a ``major event'' is very low and similar to that of the EIGs.
Therefore, ``major events'' seem to be the main mechanism creating low-mass subhaloes ($\Mhalo < 10^{10}\,\Msunh$) that host galaxies of $\AbsMg \lesssim -14$ (the magnitude range included in the Spring mock observable dataset).

\rem{
$\cz >= 1600\,\kms$ 
$\g < \g_max = 18.4$ (Spring)   (16.8 Autumn)
$D = \cz / H = \cz / (h * 100\,\kms,\Mpc^{-1})$ 
$\AbsMg = g - 5*log(D / 10\,\,\pc) = g - 15 - 5*log[ (v / \kms) / h]$
$\AbsMg < 18.4 - 15 - 5*log(2000 / 0.704) = -13.87$ Spring observable dataset
$\AbsMg < 16.8 - 15 - 5*log(2000 / 0.704) = -15.47$ Autumn observable dataset
}

\vspace{12pt}

\begin{figure*}
\begin{centering}
\includegraphics[width=14cm,trim=0mm 0mm 0mm 0, clip]{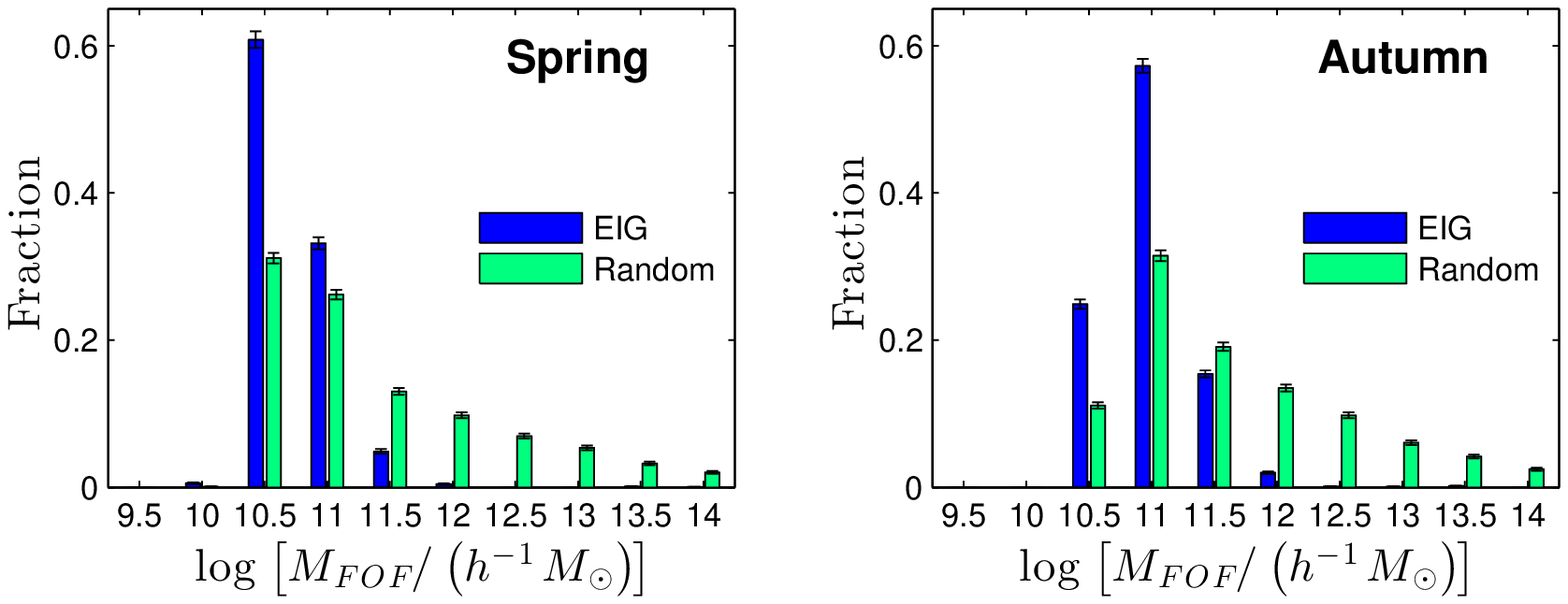}
\caption [Box160 - FOF halo mass distribution]
{ 
  Distribution of FOF halo mass, $\Mass_{FOF}$, in the Box160 mock datasets (EIG and random).\rem{ Left panel for the Spring sky region. Right panel for the Autumn region. }\label{f:Sim_Box160HaloMass}
}
\end{centering}
\end{figure*}

The PDFs of the galaxies' FOF halo mass, $\Mass_{FOF}$, were ``measured'' in Box160 and are shown in Figure \ref{f:Sim_Box160HaloMass}. The figure compares the $\Mass_{FOF}$ PDF of EIGs with that of the randomly selected galaxies for both the Spring (left panel) and Autumn (right panel).
Figure \ref{f:Sim_Box160HaloMass} shows that the $\Mass_{FOF}$ PDF includes higher masses, compared to the PDF of the subhaloes (Figure \ref{f:Sim_Mill2HaloMass}). This is expected, since each FOF halo may contain many gravitationally-bound subhaloes.
As in the case of subhaloes, the EIGs' $\Mass_{FOF}$ is less scattered than that of the random samples. The standard deviation of $\log \left[ \Mass_{FOF} / \left( \Msunh \right) \right]$ is $\sim$0.3 for the EIGs of the Spring region, compared to $\sim$1.0 for the random sample. For the Autumn region it is $\sim$0.4 for the EIGs, compared to $\sim$1.0 for the random sample.

The average $\log \left[ \Mass_{FOF} / \left( \Msunh \right) \right]$ of EIGs is smaller than that of random galaxies. In the Spring region it is 10.8 (EIGs) vs.~11.5 (random), whereas in the Autumn region it is 11.1 (EIGs) vs.~11.8 (random).
This difference may be explained by the EIGs' subhaloes being the dominant ones in their FOF haloes, as shown above. Therefore, their $\log \left[ \Mass_{FOF} / \left( \Msunh \right) \right]$ may not be significantly larger than their $\log \left[ \Mhalo / \left( \Msunh \right) \right]$. On the other hand, FOF haloes of a significant fraction of the random galaxies contain many subhaloes, which significantly increases their $\log \left[ \Mass_{FOF} / \left( \Msunh \right) \right]$.
Similarly to the case of the subhaloes, $\Mass_{FOF}$ is somewhat larger in the Autumn region, compared to the Spring region.

\subsection{History of the halo}
\label{s:SimHaloHist}

The merger trees of the haloes were analysed in the Mill2 simulation to obtain information about their Mass Accretion Histories (MAHs) and about ``major events'' in their past. Major events are defined here as either major mergers having progenitors with at least 20\% of the mass of the merged halo, or major mass-loss events in which a halo lost at least 10\% of its mass between successive simulation snapshots. Major events were counted only for the main branches of the merger trees, i.e.~major events of a satellite before it merged with a halo were disregarded.

\begin{figure*}
\begin{centering}
\includegraphics[width=14cm,trim=0mm 0mm 0mm 0, clip]{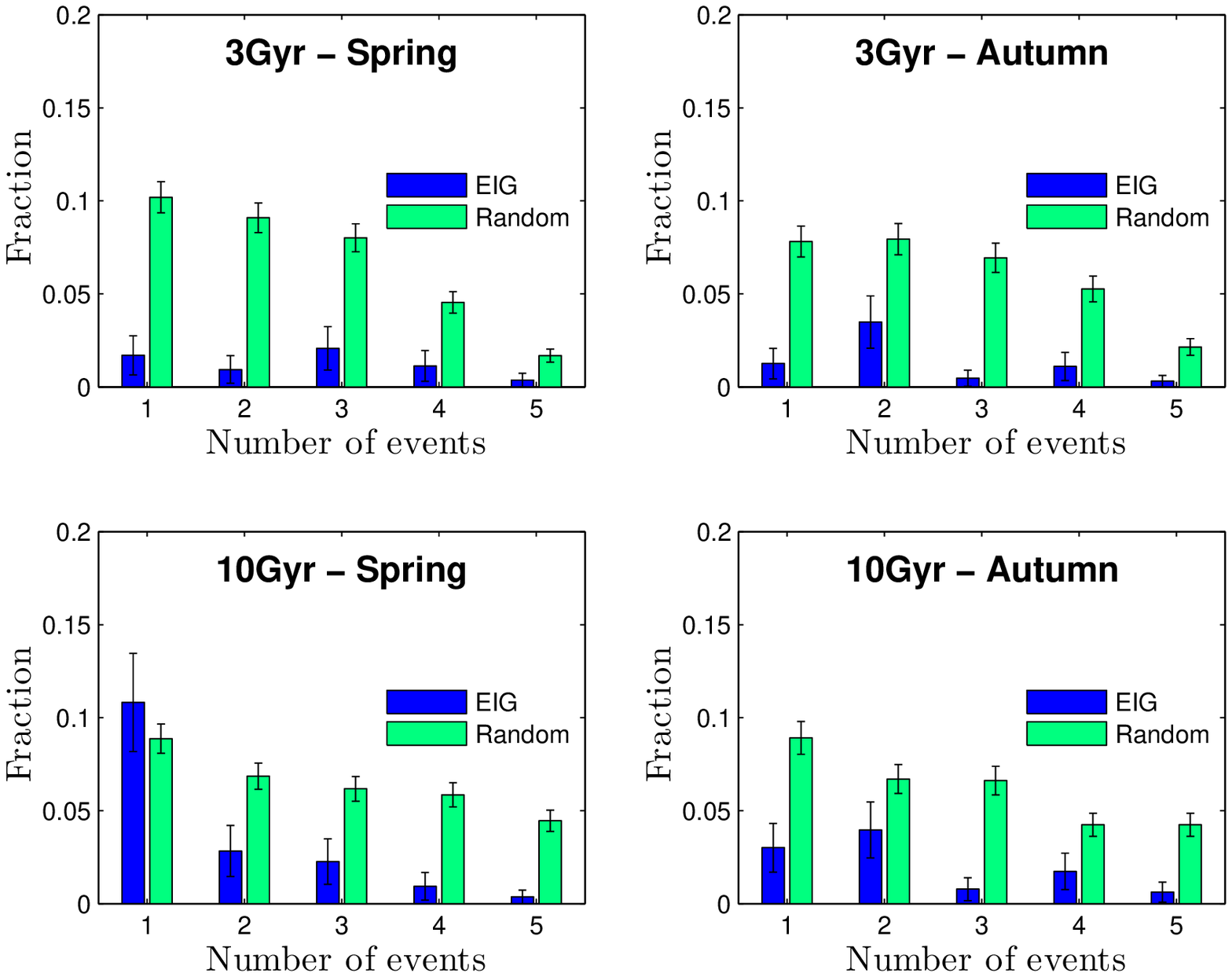}
\caption [PDF of the number of major events in the last {3\,\Gyr} and {10\,\Gyr}]
{ 
  The PDF of the number of major events in the last {$\sim$3\,\Gyr} \rem{(top panels)} and {$\sim$10\,\Gyr} \rem{(bottom panels)} for EIGs and random galaxies.
  \rem{Left panels for the Spring sky region. Right panels for the Autumn region.}
  \label{f:Sim_HaloMjrEvntNum}
}
\end{centering}
\end{figure*}

Figure \ref{f:Sim_HaloMjrEvntNum} shows the PDF of the number of major events a galaxy's subhalo went through in the past {$\sim$3\,\Gyr} (top panels) and {$\sim$10\,\Gyr} (bottom panels). The actual times ($3.16\,\Gyr$ and $9.97\,\Gyr$) are the closest simulation snapshots. As in previous figures, the left panels show the Spring region PDFs, and the right panels show the Autumn PDFs.

The probability that an EIG experienced a ``major event'' in the last {3\,\Gyr} is only $\left( 5 \pm 2 \right) \%$ for both the Spring region and the Autumn region. This, compared to $\left( 34 \pm 1 \right) \%$ for a random sample galaxy in the Spring region and $\left( 31 \pm 1 \right) \%$ for a random sample galaxy in the Autumn region. The small fraction of EIGs that did experience a ``major event'' in the last $3\,\Gyr$, is the result of both the fraction of galaxies misidentified as EIGs while in fact having unobserved close neighbours, and the possibility of a complete merger of two galaxies that left no neighbours.

\begin{figure*}
\begin{centering}
\includegraphics[width=14cm,trim=0mm 0mm 0mm 0, clip]{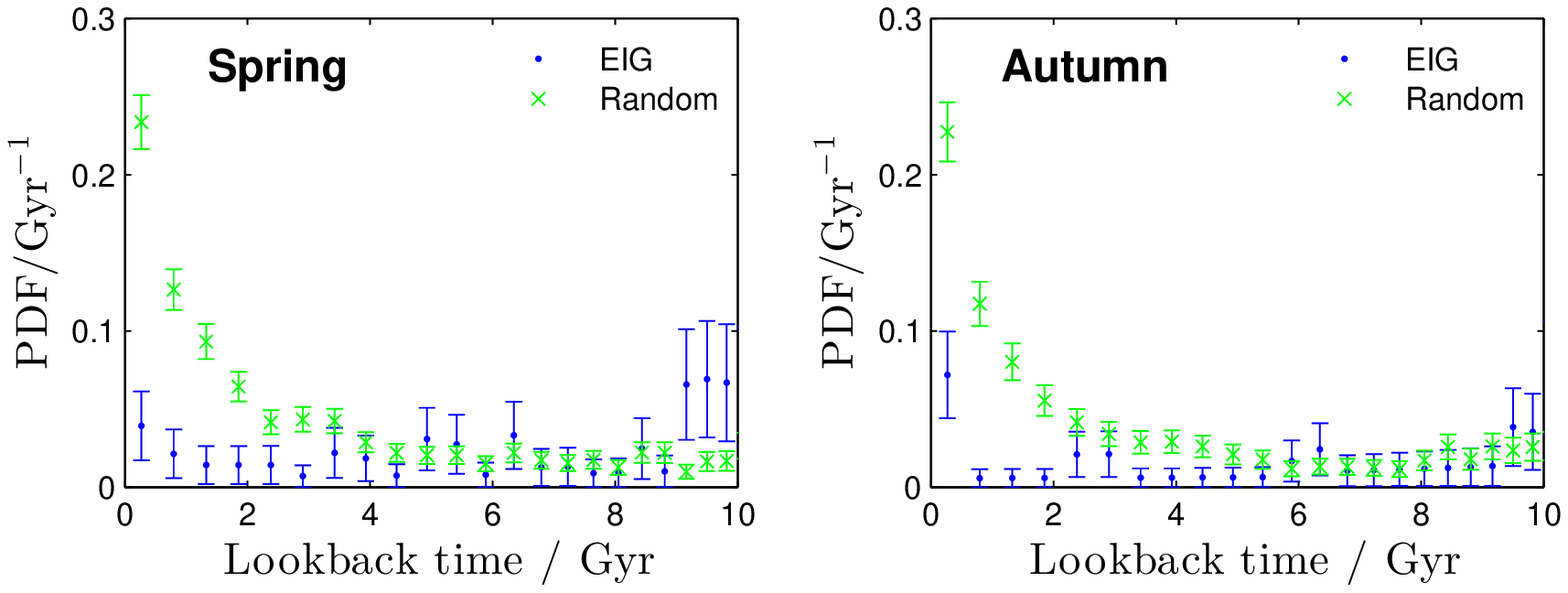}
\caption [PDF of the last major event's lookback time]
{ 
  The PDF of the last major event's lookback time (for EIGs and random).
  \rem{Left panel for the Spring sky region. Right panel for the Autumn region.}
  \label{f:Sim_HaloMjrEvntLBT}
}
\end{centering}
\end{figure*}

Figure \ref{f:Sim_HaloMjrEvntLBT} shows the probability distribution function (PDF) of the time since the last major event a galaxy's halo experienced. It appears that the PDF of the EIGs is rather similar to that of the random galaxies, except for the last $3\,\Gyr$. In the last $\sim$2\,$\Gyr$ the PDF is significantly higher for the random galaxies, and is increasing as the lookback time decreases. More than half of the random galaxies that experienced a major event in the last $3\,\Gyr$, had it in the last $1\,\Gyr$.

\rem{ LBT<=1.064Gyr from LBT<=3.165Gyr: 
 Random: $\left(60 \pm 5 \right)\%$ for Spring, $\left(62 \pm 5 \right) \%$ for Autumn
 : $\left(55 \pm 30 \right) \%$ for Spring, $\left(59 \pm 27 \right) \%$ for Autumn
}

\begin{figure*}
\begin{centering}
\includegraphics[width=14cm,trim=0mm 0mm 0mm 0, clip]{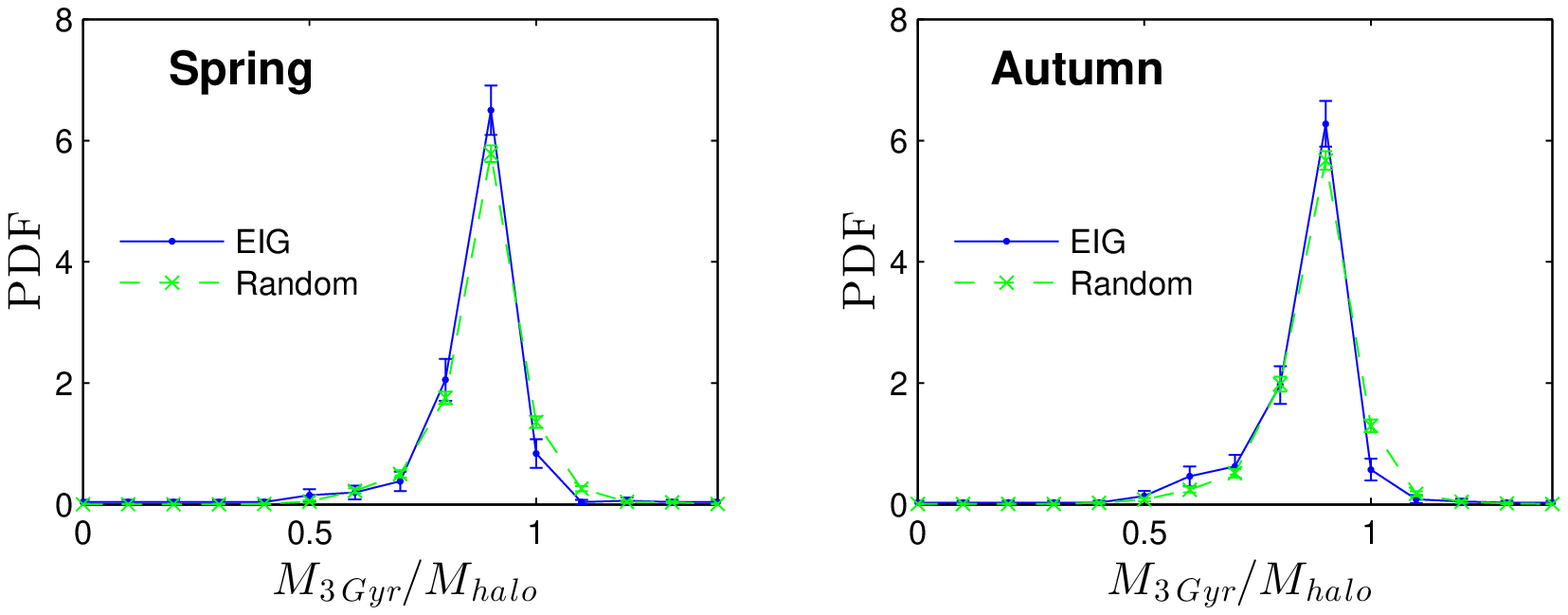}
\caption [PDF of the fraction of halo mass, accreted by {3\,\Gyr} lookback time]
{  PDF of the fraction of halo mass, accreted by {3\,\Gyr} lookback time (for EIGs and random).
  \rem{Left panel for the Spring sky region. Right panel for the Autumn region.}
  \label{f:Sim_HaloMassFrac3Gyr}
}
\end{centering}
\end{figure*}

\begin{figure*}
\begin{centering}
\includegraphics[width=14cm,trim=0mm 0mm 0mm 0, clip]{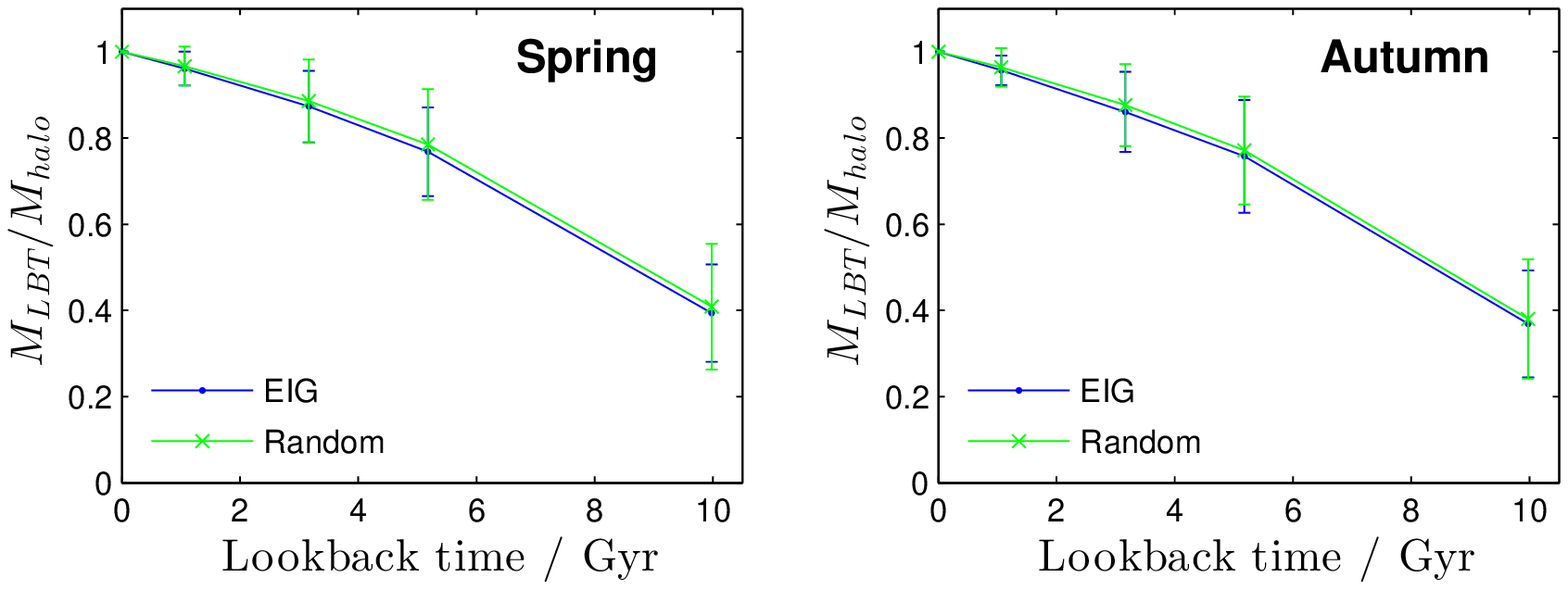}
\caption [Fraction of halo mass, already accreted, as function of the lookback time]
{
  Fraction of halo mass, already accreted, as function of the lookback time (for EIGs and random galaxies that did not experience major events in the last $10\,\Gyr$). The bars indicate the standard deviation of this fraction in the sample (rather than the measurement uncertainties).
  \rem{Left panel for the Spring sky region. Right panel for the Autumn region.}
  \label{f:Sim_HaloAvgMassFrac}
}
\end{centering}
\end{figure*}

The mass accretion history (MAH) of the haloes that did not experience a major event in the last $10\,\Gyr$ was analysed. The results are shown in Figures \ref{f:Sim_HaloMassFrac3Gyr} and \ref{f:Sim_HaloAvgMassFrac}. In both figures the MAH is measured as the fraction of the current halo mass which the halo accumulated at a particular lookback time (LBT), $\Mass_{LBT} / \Mhalo$. 
Figure \ref{f:Sim_HaloMassFrac3Gyr} shows the PDF of this fraction at $LBT = 3\,\Gyr$ and indicates that the PDFs of EIGs and random galaxies are quite similar.

Figure \ref{f:Sim_HaloAvgMassFrac} shows the average and standard deviation (indicated by bars) of $\Mass_{LBT} / \Mhalo$ of haloes that did not experience major events in the last $10\,\Gyr$. Again, it is evident that the MAHs of the EIGs are very similar to those of the random galaxies, and that there is no significant difference between the Spring and the Autumn samples.

These results indicate that mass accretion (relative to the current halo mass, $\Mhalo$) is affected by the halo's environment, mainly through strong interactions with its neighbours. As long as the halo does not experience major events, its MAH does not depend significantly on its environment.

\subsection{Neighbourhood galaxies}
\label{s:SimNgbrhdGals}

\begin{figure*}
\begin{centering}
\includegraphics[width=14cm,trim=0mm 0mm 0mm 0, clip]{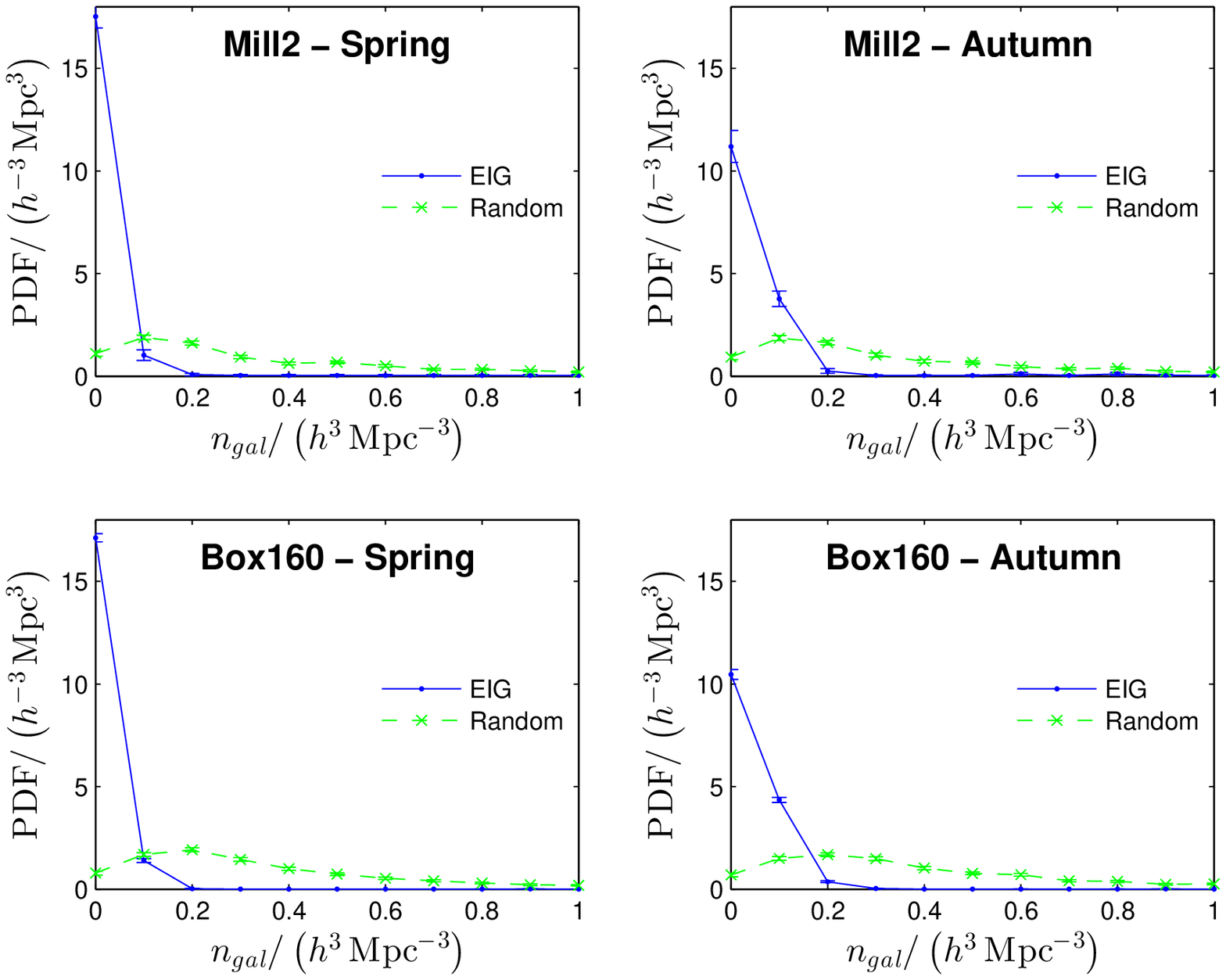}
\caption [PDF of $n_{gal}$ ]
{
   PDF of the number density of neighbouring galaxies, $n_{gal}$, with $\AbsMg < -15$ in a $3\,\Mpch$ radius sphere around the EIGs or random galaxies. \rem{The left panels are for the Spring region, and the right panels for the Autumn region. The upper panels are calculated from the Mill2 simulation, and the bottom panels from the Box160 simulation.}\label{f:Sim_n_gal_3Mpch_min15}
}
\end{centering}
\end{figure*}

The number density of neighbouring galaxies, $n_{gal}$, around EIGs and around the random galaxies was analysed in both Box160 and Mill2. Figure \ref{f:Sim_n_gal_3Mpch_min15} shows the results for spheres of radius $3\,\Mpch$ around the galaxies and for neighbour galaxies brighter than $\AbsMg = -15$. Each panel compares the $n_{gal}$ PDF of EIGs to that of the random galaxies. The left panels show results for Spring and the right panels for Autumn. The upper panels were calculated from Mill2 data and the lower ones from Box160 data.

It is evident that the EIGs are located in significantly underdense environments compared to the general population (the random samples). It is also evident that the Spring EIGs tend to reside in lower density regions, compared to the Autumn EIGs.

The results calculated from Mill2 and Box160 agree quite well, although the difference between the Mill2 and Box160 PDF points are somewhat larger than the calculated 95\% ($2\sigma$) confidence level errors shown in the figure.

The fraction of Spring EIGs residing in environments with $n_{gal} < 0.1\,\NumCntMpch$ is $0.98 \pm 0.01$ (for $\AbsMg < -15$ neighbours in a $3\,\Mpch$ radius sphere). The fraction of Autumn EIGs in such environments is $0.87 \pm 0.01$. This, vs.~0.11--0.15 of the random galaxies that reside in such $n_{gal} < 0.1\,\NumCntMpch$ environments.
The fraction of Spring EIGs that reside in much denser environments, with $n_{gal} > 0.5\,\NumCntMpch$, is $0.02 \pm 0.01$ (Mill2) or $0.002 \pm 0.001$ (Box160). For Autumn EIGs, the fraction is $0.04 \pm 0.02$ (Mill2) or $0.004 \pm 0.001$ (Box160). This, vs.~0.32--0.40 for the random galaxies.
The ``tail'' of the PDF of the random galaxies continues far beyond {1\,\NumCntMpch}. In fact, a significant fraction, 0.21--0.22 (Mill2) or 0.13 (Box160), of the random galaxies reside in $n_{gal} > 1.0\,\NumCntMpch$ environments, as could be expected given the clustering of galaxies.

\vspace{12pt}

\begin{figure*}
\begin{centering}
\includegraphics[width=14cm,trim=0mm 0mm 0mm 0, clip]{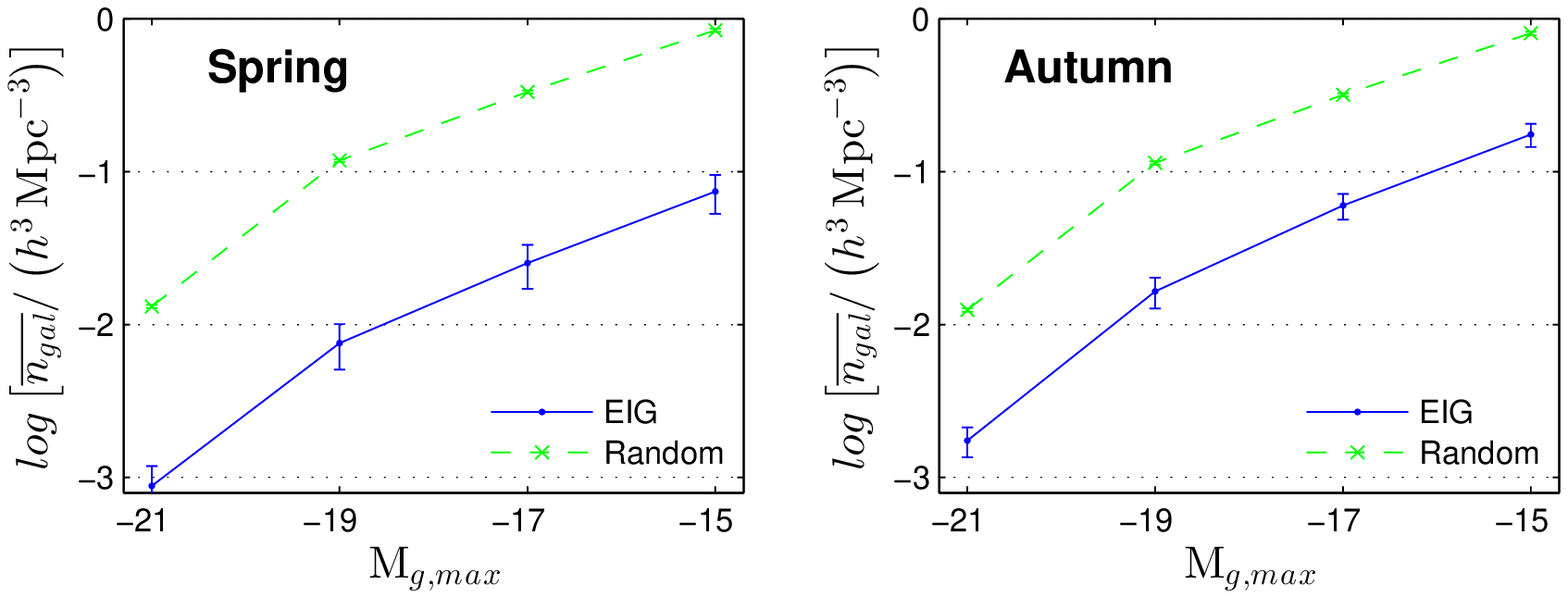}
\caption [Average number density of galaxies, $\overline{n_{gal}}$, as function of limiting neighbour absolute magnitude]
{
   Average number density of neighbouring galaxies, $\overline{n_{gal}}$, in a $3\,\Mpch$ radius sphere around the EIGs or random galaxies, as function of limiting absolute magnitude, $\mbox{M}_{g,max}$ (calculated using Mill2). 
   \rem{The left panels are for the Spring region, and the right panels for the Autumn region.}
   \label{f:Sim_n_gal_vs_mag}
}
\end{centering}
\end{figure*}

\begin{figure*}
\begin{centering}
\includegraphics[width=14cm,trim=0mm 0mm 0mm 0, clip]{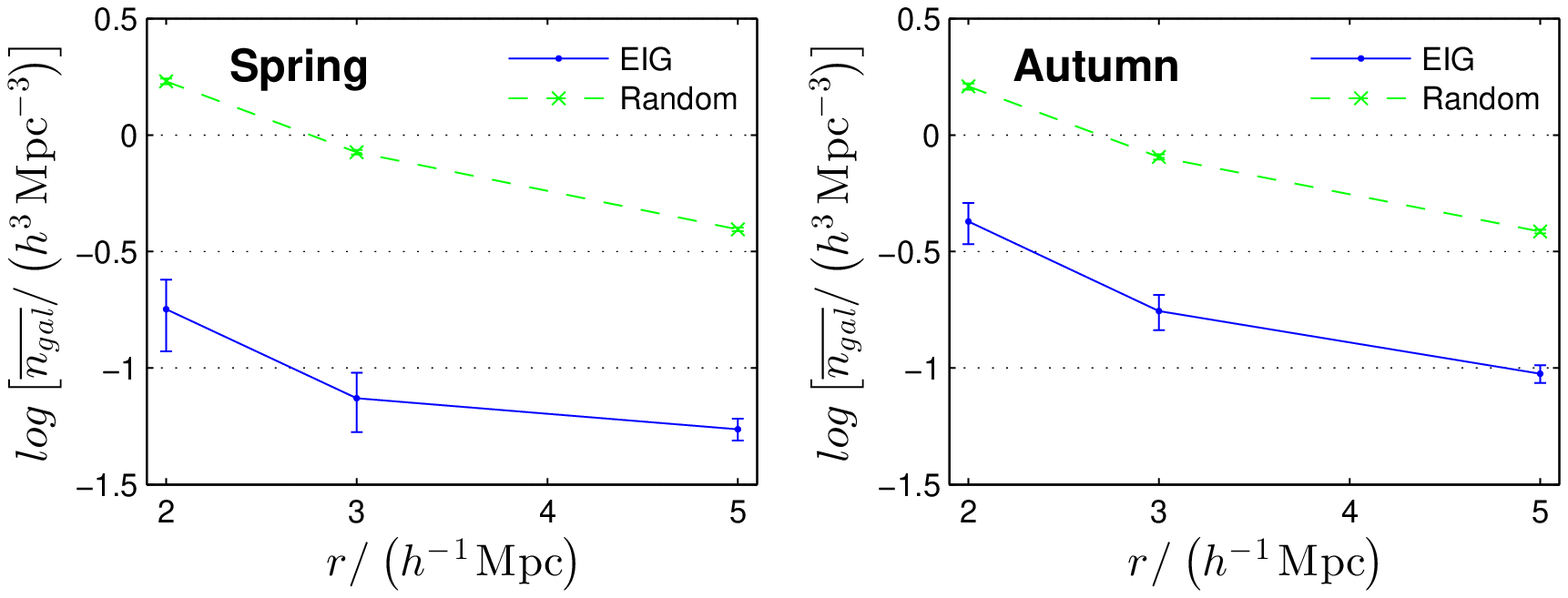}
\caption [Average number density of galaxies, $\overline{n_{gal}}$, as function of distance]
{
   Average number density of neighbouring galaxies, $\overline{n_{gal}}$, with $\AbsMg < -15$ as function of the sphere's radius around the EIGs or random galaxies, $r$, in which the density is averaged (calculated using Mill2). 
   \rem{The left panels are for the Spring region, and the right panels for the Autumn region.}
   \label{f:Sim_n_gal_vs_dist}
}
\end{centering}
\end{figure*}

The average number density of neighbouring galaxies, $\overline{n_{gal}}$, was also analysed. Figure \ref{f:Sim_n_gal_vs_mag} shows $\overline{n_{gal}}$ as function of the limiting absolute magnitude, $\mbox{M}_{g,max}$, of the neighbouring galaxies counted in a $3\,\Mpch$ radius sphere. Figure \ref{f:Sim_n_gal_vs_dist} shows $\overline{n_{gal}}$ as function of the sphere radius, $r$, for $\mbox{M}_{g,max} = -15$.\rem{ Both figures show results from the Mill2 simulation for EIGs and random galaxies, and for the Spring and Autumn sky regions.}
Both figures show that $\overline{n_{gal}}$ of the Spring EIGs is about one order of magnitude smaller than that of random samples in the magnitude limit range $-21 < \mbox{M}_{g,max} < -15$ (Figure \ref{f:Sim_n_gal_vs_mag}) and in the sphere radii range $2 < r < 5\,\Mpch$ (Figure \ref{f:Sim_n_gal_vs_dist}). For the Autumn region, $\overline{n_{gal}}$ of the EIGs is 4 to 8 times smaller than that of the random samples in these ranges. The dependence of the EIGs' $\overline{n_{gal}}$ on the limiting magnitude, $\mbox{M}_{g,max}$, and on the sphere radius, $r$, is quite similar to that of random galaxies.

\subsection{Neighbourhood haloes}
\label{s:SimNgbrhdHalos}

\begin{figure*}
\begin{centering}
\includegraphics[width=14cm,trim=0mm 0mm 0mm 0, clip]{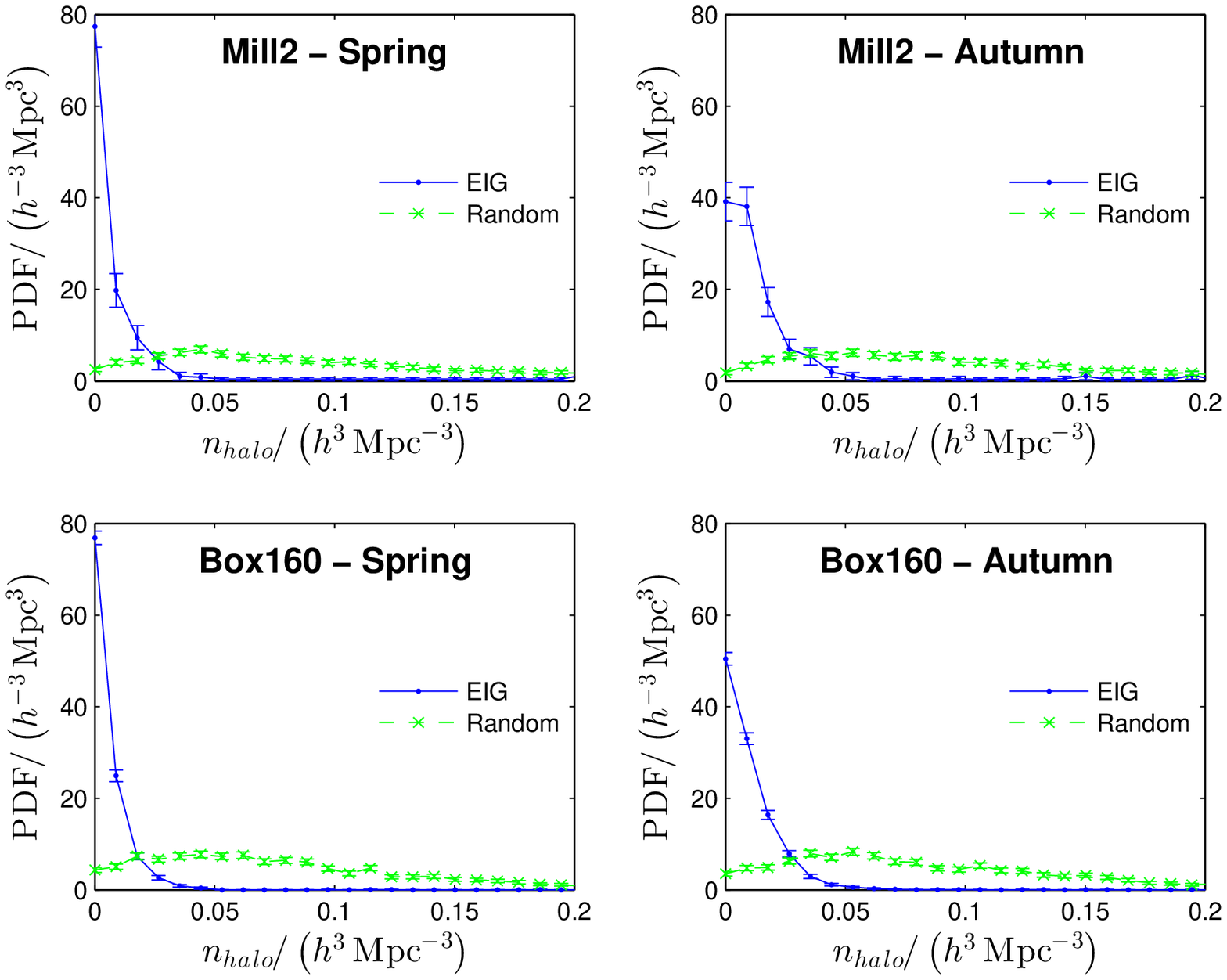}
\caption [PDF of $n_{halo}$ ]
{
   PDF of the number density of neighbouring haloes, $n_{halo}$, with $\Mhalo > 10^{11}\,\Msunh$ in a $3\,\Mpch$ radius sphere around the EIGs or random galaxies. 
   \rem{The left panels are for the Spring region, and the right panels for the Autumn region. The upper panels are calculated from the Mill2 simulation, and the bottom panels from the Box160 simulation.}
   \label{f:Sim_n_halo_3Mpch_min1e10}
}
\end{centering}
\end{figure*}

The number density of neighbouring haloes, $n_{halo}$, around EIGs and around the random galaxies was analysed for both the FOF haloes of Box160 and the gravitationally-bound subhaloes of Mill2. The results are similar to those of the $n_{gal}$ analysis. Figure \ref{f:Sim_n_halo_3Mpch_min1e10} shows the results for neighbouring haloes with mass $\Mhalo > 10^{11}\,\Msunh$ in spheres of radius $3\,\Mpch$ around the galaxies.

As in the case of $n_{gal}$, Figure \ref{f:Sim_n_halo_3Mpch_min1e10} shows that EIGs populate environments significantly less dense than those of the general population (random samples), and that Spring EIGs tend to reside in lower density regions compared to the Autumn EIGs.
The results calculated from Mill2 and Box160 agree. The difference between the Mill2 and Box160 PDFs is mostly within the plotted 95\% confidence level errors.
The fraction of Spring EIGs that reside in environments with $n_{halo} < 0.03\,\NumCntMpch$ is $0.98 \pm 0.01$ (for $\Mhalo > 10^{11}\,\Msunh$ neighbours in a $3\,\Mpch$ radius sphere). The fraction of Autumn EIGs that reside in such environments is $0.89 \pm 0.02$ (Mill2) or $0.95 \pm 0.01$ (Box160). This, vs.~0.13--0.21 for the random samples.

The fraction of Spring EIGs that reside in environments with $n_{halo} > 0.1\,\NumCntMpch$ is $0.02 \pm 0.01$ (Mill2) or $0.002 \pm 0.001$ (Box160), whereas for Autumn EIGs it is $0.04 \pm 0.02$ (Mill2) or $0.003 \pm 0.001$ (Box160). This, vs.~0.32--0.48 for the random samples.
The ``tail'' of the PDF of the random galaxies continues beyond 0.2\,\NumCntMpch. A fraction of $0.23 \pm 0.01$ (Mill2) or $0.09 \pm 0.01$ (Box160) of the random galaxies reside in $n_{halo} > 0.2\,\NumCntMpch$ environments.

\vspace{12pt}

\begin{figure*}
\begin{centering}
\includegraphics[width=14cm,trim=0mm 0mm 0mm 0, clip]{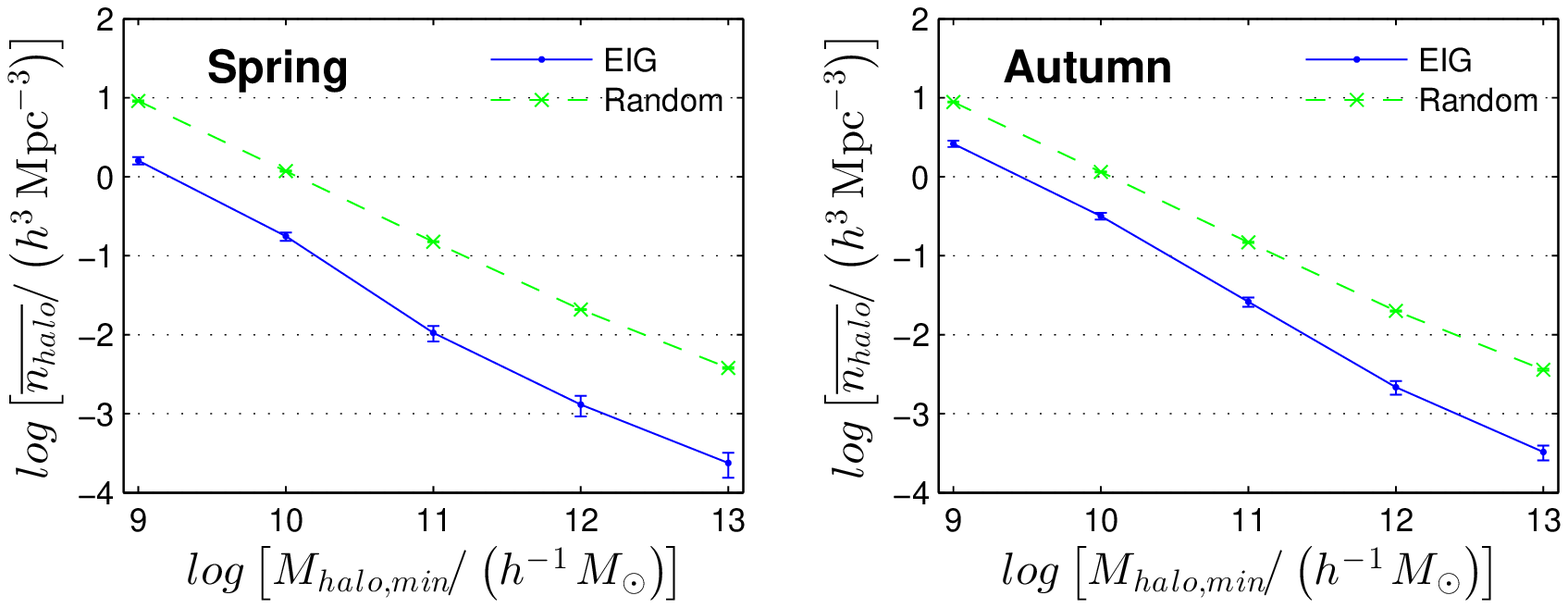}
\caption [Average number density of haloes, $\overline{n_{halo}}$, as function of limiting halo mass]
{
   Average number density of neighbouring haloes, $\overline{n_{halo}}$, in a $3\,\Mpch$ radius sphere around the EIGs or random galaxies, as function of minimum halo mass, $\Mass_{halo,min}$ (calculated using Mill2). 
   \rem{The left panels are for the Spring region, and the right panels for the Autumn region.}
   \label{f:Sim_n_halo_vs_MassLimit}
}
\end{centering}
\end{figure*}

\begin{figure*}
\begin{centering}
\includegraphics[width=14cm,trim=0mm 0mm 0mm 0, clip]{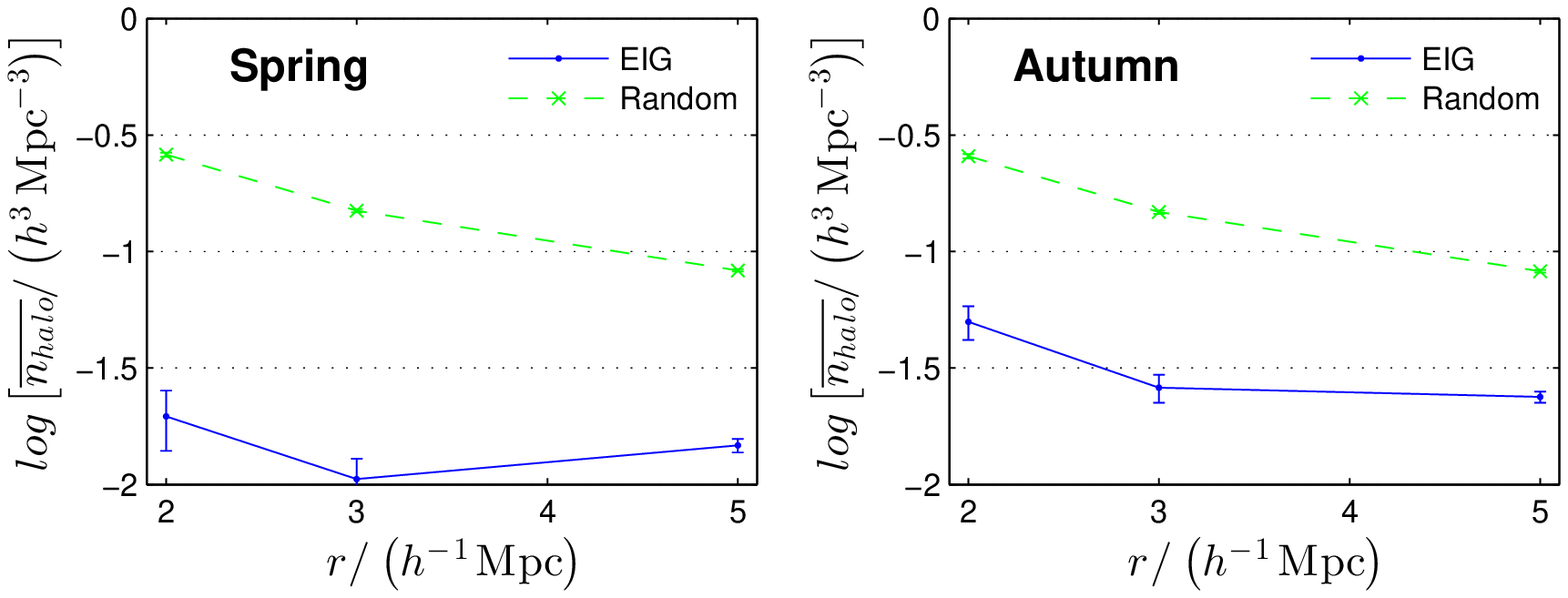}
\caption [Average number density of haloes, $\overline{n_{halo}}$, as function of distance]
{
   Average number density of neighbouring haloes, $\overline{n_{halo}}$, with $\Mhalo > 10^{11}\,\Msunh$ as function of the sphere's radius around the EIGs or random galaxies, $r$, in which the density is averaged (calculated using Mill2). 
   \rem{The left panels are for the Spring region, and the right panels for the Autumn region.}
   \label{f:Sim_n_halo_vs_dist}
}
\end{centering}
\end{figure*}

The average number density of neighbouring haloes, $\overline{n_{halo}}$ was also analysed. Figure \ref{f:Sim_n_halo_vs_MassLimit} shows $\overline{n_{halo}}$ as function of the limiting halo mass, $\Mass_{halo,min}$, of the neighbouring haloes counted in a $3\,\Mpch$ radius sphere. Figure \ref{f:Sim_n_halo_vs_dist} shows $\overline{n_{halo}}$ as function of the sphere radius, $r$, for $\Mhalo > 10^{11}\,\Msunh$ neighbours.
Both figures show that $\overline{n_{halo}}$ of the Spring EIGs is about an order of magnitude smaller than that of random samples in the range $9 < \log \left[ \Mass_{halo,min} / \left( \Msunh \right) \right] < 13$ (Figure \ref{f:Sim_n_halo_vs_MassLimit}) and $2 < r < 5\,\Mpch$ (Figure \ref{f:Sim_n_halo_vs_dist}). For the Autumn region, $\overline{n_{halo}}$ of the EIGs is 3 to 10 times smaller than that of the random samples in these ranges.
The dependence of the EIGs' $\overline{n_{halo}}$ on the limiting halo mass, $\Mass_{halo,min}$, and on the sphere radius, $r$, is quite similar to that of the random galaxies' $\overline{n_{halo}}$, with $\overline{n_{halo}}$ of the EIG\rem{ dropping slightly faster with an increase in $\Mass_{halo,min}$ and slightly slower with an increase in $r$}.

\subsection{Neighbourhood mass density}
\label{s:SimNgbrhdMassDens}

\begin{figure*}
\begin{centering}
\includegraphics[width=14cm,trim=0mm 0mm 0mm 0, clip]{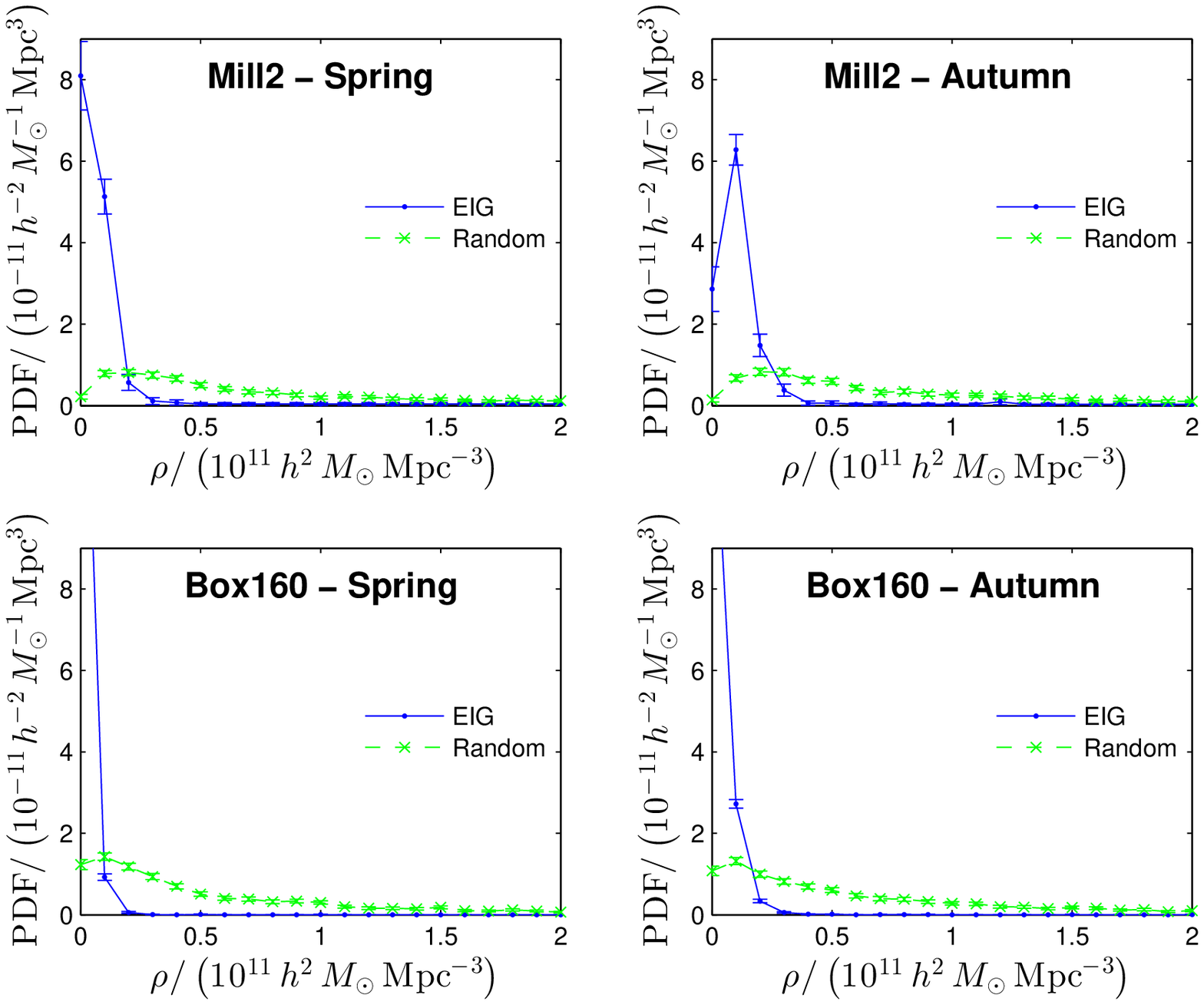}
\caption [PDF of neighbourhood density ]
{
   PDF of the mass density of neighbouring haloes, $\density$, in a $3\,\Mpch$ radius sphere around the EIGs or random galaxies. 
   \rem{The left panels are for the Spring region, and the right panels for the Autumn region. The upper panels are calculated from the Mill2 simulation, and the bottom panels from the Box160 simulation.}
   \label{f:SimMassDensity_3Mpch}
}
\end{centering}
\end{figure*}

The mass density of neighbouring haloes, $\density$, was analysed in both the Mill2 and Box160 simulations (each with its limiting {\Mhalo}) within three spheres of radius $r = 2,\: 3, \: 5\,\Mpch$. Figure \ref{f:SimMassDensity_3Mpch} show the probability distribution functions (PDFs) of $\density$, calculated for $r = 3\,\Mpch$.

Figure \ref{f:SimMassDensity_3Mpch} shows that EIGs populate the low-$\density$ environments, while random galaxies populate both low and high $\density$ environments. The Spring sky region EIGs (left panels) are generally located in lower $\density$ environments compared to the Autumn EIGs (right panels). This is a result of the higher completeness of the Spring dataset, given the assumption of an average composition of the large-scale structure (as discussed in section \ref{s:completenessFunc}).

All PDFs calculated using Box160 (bottom panels) are somewhat shifted to lower $\density$ compared to their Mill2 (top panels) parallels. This shift can be attributed mainly to the higher neighbouring halo mass limit of Box160, $\Mhalo \geq 1.8 \cdot 10^{10} \Msunh$, vs.~$\Mhalo \geq 10^{9}\,\Msunh$ for Mill2.

The fraction of Spring EIGs that reside in $\density < 0.25 \cdot 10^{11}\,\densityMsunMpch$ environments (averaged on a $3\,\Mpch$ radius sphere) is $0.98 \pm 0.01$, whereas for Autumn EIGs it is $0.91 \pm 0.02$ (Mill2) or $0.988 \pm 0.003$ (Box160). This, vs.~0.16--0.32 for the random samples.
The fraction of Spring EIGs that reside in $\density > 10^{11}\,\densityMsunMpch$ environments is $0.02 \pm 0.01$ (Mill2) or $0.001 \pm 0.001$ (Box160), while for Autumn EIGs it is $0.04 \pm 0.02$ (Mill2) or $0.002 \pm 0.001$ (Box160). This, vs.~0.30--0.50 for the random samples.
The ``tail'' of the PDF of the random galaxies continues beyond $2 \cdot 10^{11}\,\densityMsunMpch$. A fraction of $0.34 \pm 0.01$ (Mill2) or $0.16 \pm 0.01$ (Box160) of the random galaxies reside in $\density > 2 \cdot 10^{11}\,\densityMsunMpch$ environments.

\vspace{12pt}

\begin{figure*}
\begin{centering}
\includegraphics[width=14cm,trim=0mm 0mm 0mm 0, clip]{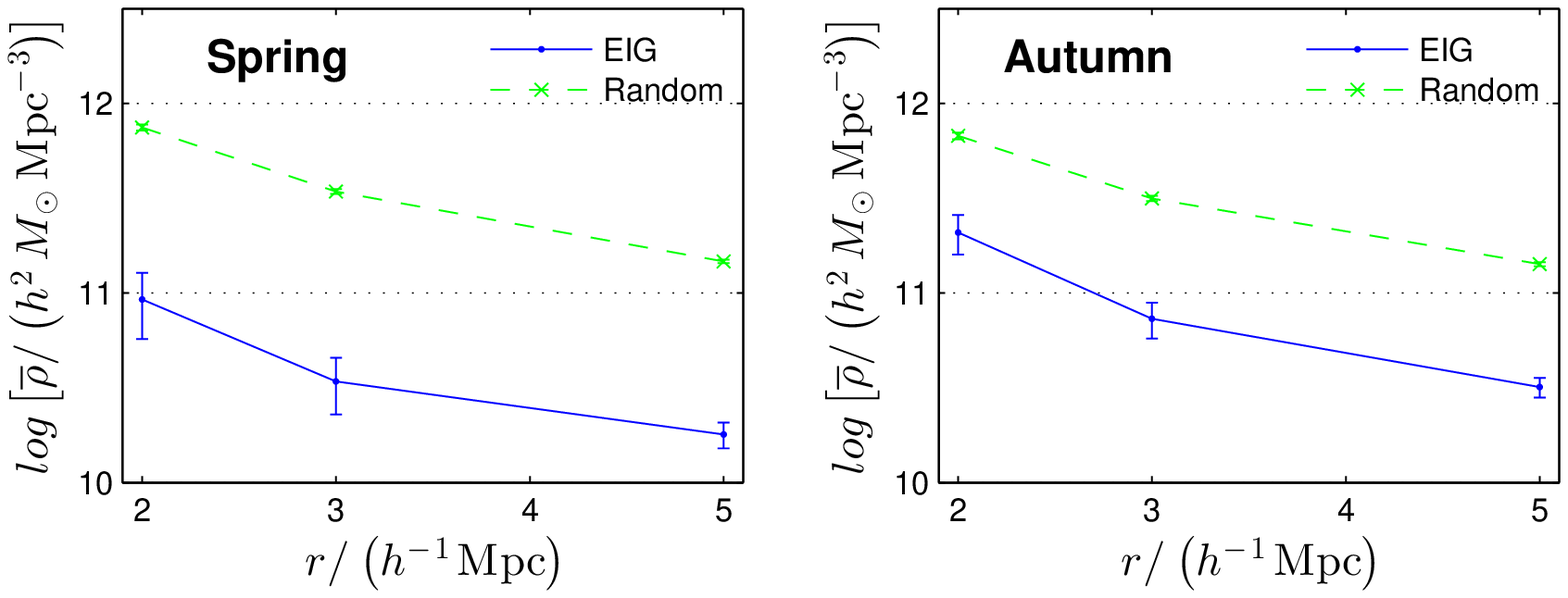}
\caption [Average mass density of haloes, $\overline{n_{halo}}$, as function of distance]
{
   Average mass density of neighbouring haloes, $\overline{\density}$, as function of the sphere's radius around the EIGs or random galaxies, $r$, in which the density is averaged (calculated using Mill2). 
   \rem{The left panels are for the Spring region, and the right panels for the Autumn region.}
   \label{f:SimMassDensity_vs_dist}
}
\end{centering}
\end{figure*}

The average mass density of neighbouring haloes, $\overline{\density}$, was also analysed. Figure \ref{f:SimMassDensity_vs_dist} shows the Mill2 results as function of the sphere radius, $r$.
The figure show that $\overline{\density}$ for Spring EIGs is about one order of magnitude smaller than that of the random samples in spheres with radius in the range $2 < r < 5\,\Mpch$. For the Autumn region, $\overline{\density}$ of the EIGs is 3 to 4 times smaller than that of the random samples in this range. The dependence of the EIGs' $\overline{\density}$ on $r$ is similar to that of random galaxies.

\subsection{Tidal acceleration}
\label{s:SimTide}

\begin{figure*}
\begin{centering}
\includegraphics[width=14cm,trim=0mm 0mm 0mm 0, clip]{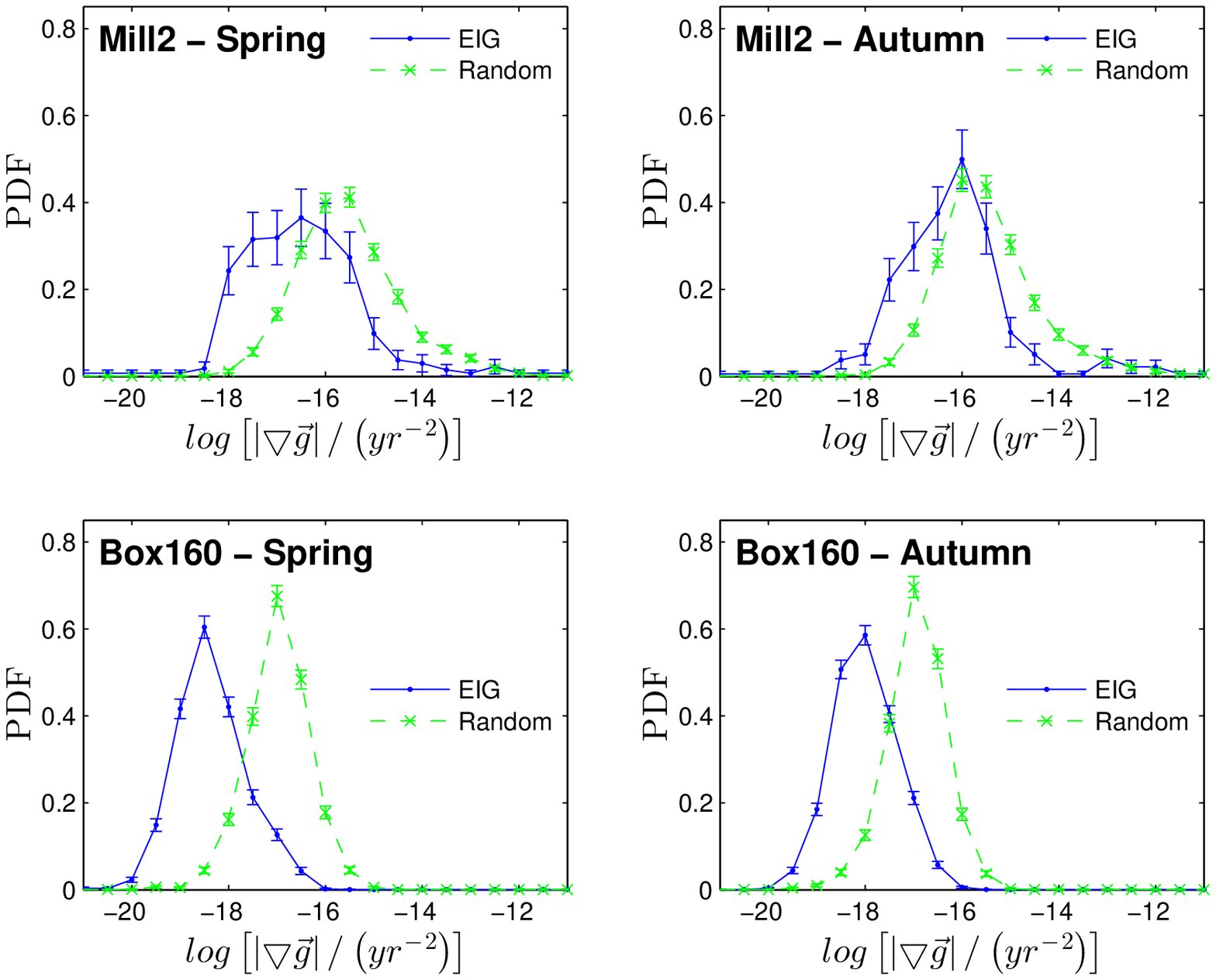}
\caption [PDF of tidal force ]
{
   PDF of the tidal acceleration logarithm, $log \left[ \left| \bigtriangledown \vec{g} \right| / \left( yr^{-2} \right) \right]$, on EIGs or random galaxies. 
   \rem{The left panels are for the Spring region, and the right panels for the Autumn region. The upper panels are calculated from the Mill2 simulation, and the bottom panels from the Box160 simulation.}
   \label{f:SimTidalForce}
}
\end{centering}
\end{figure*}

The tidal acceleration is defined here as the difference in gravitational acceleration exerted by neighbouring haloes, $\vec{g}$, per unit displacement, i.e. the divergence of this gravitational acceleration, $\bigtriangledown \vec{g}$. The total tidal acceleration, $\bigtriangledown \vec{g}$, that haloes within {5\,\Mpch} exert on a galaxy was calculated for each mock EIG and random galaxy, with the approximation that the haloes' masses are concentrated at their centre. Since $\bigtriangledown \vec{g}$ varies by orders of magnitude, the PDFs were calculated for its logarithm, $\log \left[ \left| \bigtriangledown \vec{g} \right| / \left( yr^{-2} \right) \right]$. The results are shown in Figure \ref{f:SimTidalForce}.

Both simulations show that the average tidal acceleration, $\bigtriangledown \vec{g}$, of EIGs is about an order of magnitude smaller than that of random galaxies. The difference in $log \left[ \left| \bigtriangledown \vec{g} \right| / \left( yr^{-2} \right) \right]$ is larger for the Spring ($-1.0$ according to Mill2, and $-1.4$ according to Box160) compared to the Autumn ($-0.8$ according to Mill2, and $-1.1$ according to Box160). This again is the result of the lower completeness estimated for the Autumn region (see section \ref{s:completenessFunc}).

There is a discrepancy between the results of the Box160 simulation and those of Mill2. The average $log \left[ \left| \bigtriangledown \vec{g} \right| / \left( yr^{-2} \right) \right]$ of Box160 is lower than that of Mill2 by $-1.5$ for the random galaxies, and $-1.8$ for the EIGs. This discrepancy can be attributed mainly to the difference in the halo mass limit of the two databases ($\Mhalo \geq 1.8 \cdot 10^{10} \Msunh$ for Box160, and $\Mhalo \geq 10^{9}\,\Msunh$ for Mill2). However, this could also be a result of the Box160 calculation being based on FOF haloes and the Mill2 on gravitationally-bound haloes, or to the different cosmological parameters applied in the two simulations. If the halo mass limit has a strong effect, the actual $\left| \bigtriangledown \vec{g} \right|$ is expected to be even higher than the Mill2 curve indicates.

The values of the tidal acceleration, $\bigtriangledown \vec{g}$, can be interpreted as indicating the typical time of induced change in the shape of objects by external interactions. For example, consider two nearby gas clouds with no relative motion in a constant $\bigtriangledown \vec{g}$ tidal field. If the gravitation between these two clouds is negligible, the tidal field will drive the clouds apart, doubling their distance within $t \cong 1.32 \left| \bigtriangledown \vec{g} \right|^{-0.5}$. In general, the time in which a constant tidal acceleration field, $\bigtriangledown \vec{g}$, changes the shape of objects (or the distance between objects) is proportional to $\left| \bigtriangledown \vec{g} \right|^{-0.5}$.

The results of the tidal acceleration analysis, therefore, indicate that the time of induced change is $\sim$3 times larger, on average, for EIGs in comparison to random galaxies. This may have a significant effect on the star formation rate (SFR), if this time of scale change induced by neighbouring haloes is in the order of magnitude of typical shape change times induced by internal sources.

The discrepancy between the $\left| \bigtriangledown \vec{g} \right|$, calculated from Box160 data, and that calculated from Mill2, translates to a factor of 5--8 in this typical time of induced shape change (longer time for Box160). The actual shape change times are expected to be even shorter than those calculated using Mill2, if calculated with a mass limit lower than $10^{9}\,\Msunh$. A possible overestimate of the low mass satellite haloes in the simulations (the missing satellite problem) might also affect the results, in which case the actual $\left| \bigtriangledown \vec{g} \right|$ will be smaller than indicated by Mill2 and the typical time of induced shape change will be longer. This might have an effect on SFR of EIGs and random galaxies, where EIGs might be affected differently than random galaxies.

%% file: Chapters/7_Conclusions.tex
\section{Conclusions}
\label{ch:Conclusions}

The simple isolation criterion (having no known neighbours within {3\,\Mpch}), when applied to redshifts in the range $2000<\cz<7000\,\kms$ using NED and ALFALFA data, provides a sample of galaxies that are extremely isolated, compared to the general population.

\vspace{12pt}

Using cosmological simulations, we confirmed that the EIG-1 and EIG-2 subsamples are a subset of galaxies significantly more isolated than the general galaxy population. Apart from the low density regions in which they reside, EIGs  are characterized by normal mass haloes, which have evolved gradually with little or no major mergers or major mass-loss events. As a result of their low-density environments, the tidal acceleration exerted on EIGs is typically about one order of magnitude lower than the average tidal acceleration exerted on the general population of galaxies.

The level of contamination in the sample, i.e. the fraction of EIGs which are not in extremely isolated environments or which experienced strong interactions in the last {3\,\Gyr}, was found to be {5\%--10\%}. The Spring EIGs seem to be more isolated than the Autumn EIGs.

We have defined a major event as either a major merger (where the progenitors are at least 20\% of the mass of the merged halo) or a major mass-loss event (where a halo lost at least 10\% of its mass between successive simulation snapshots).
We have found that this definition includes almost all interactions between haloes that are strong enough to significantly alter their mass accretion histories (MAHs). Haloes that did not experience major events in the last {10\,\Gyr} accreted matter very similarly (when comparing after normalization with the current halo mass).

We have found that almost all low-mass haloes ($\Mhalo < 10^{10}\,\Msunh$) that produced enough stars to be included in redshift surveys of the local Universe ($\AbsMg \lesssim -14$) experienced major events in their past. Therefore, Mill2 simulation results predict that samples of low-dark-mass galaxies are biased in the sense that they hardly include any galaxies that did not experience major events in their histories. 
We have found that EIGs are very unlikely to reside in such low-mass haloes, probably as a result of the fact that they hardly experience major events.

\vspace{12pt}

ALFALFA data are extremely useful in improving the sample's isolation level and in eliminating false positives, due to the redshift data it provides for low-luminosity galaxies with high HI masses. The EIG-1 subsample galaxies (EIGs which passed the isolation criterion with ALFALFA data, as well as with NED data) are significantly more isolated than the EIG-2 galaxies.

\vspace{12pt}

The properties of the EIG samples, derived from observations, will be described in a forthcoming paper (Spector \& Brosch in preparation).

%% file: Chapters/acknowledgments.tex
\section*{Acknowledgements}

{\rem{ALFALFA}}
We are grateful to Martha Haynes, Riccardo Giovanelli and the entire ALFALFA team for providing an unequalled HI data set.
{\rem{Box160}}
We are grateful to Yehuda Hoffman, Stefan Gottl\"{o}ber and Ofer Metuki for providing the Box 160 simulation dataset.
{\rem{Referee}}
\markChange{We are grateful to the anonymous referee for some constructive remarks that improved the paper.}

{\rem{NED}}
This research has made use of the NASA/IPAC Extragalactic Database (NED) which is operated by the Jet Propulsion Laboratory, California Institute of Technology, under contract with the National Aeronautics and Space Administration.
{\rem{Mill2}}
The Millennium-II Simulation databases used in this work and the web application providing online access to them were constructed as part of the activities of the German Astrophysical Virtual Observatory (GAVO).
{\rem{SDSS}}
Funding for SDSS-III has been provided by the Alfred P. Sloan Foundation, the Participating Institutions, the National Science Foundation, and the U.S. Department of Energy Office of Science. The SDSS-III web site is http://www.sdss3.org/.
\rem{this is part of the Official SDSS-III Acknowledgement, which can be found on: http://www.sdss3.org/collaboration/boiler-plate.php}

%% file: Chapters/appB_IndividualGalaxies.tex
\section{\\EIG Specific Data}
\label{App:EIGdata}

\markChange
{
This appendix contains general notes for some of the EIGs. It also discusses galaxies which were originally identified as isolated but were eventually not included in the sample.
}

\subsection{Notes for specific EIGs}
\label{App:EIGdata.Notes}

\subsubsection*{EIG 1s-05}

\markChange
{
No optical counterpart could be identified for EIG 1s-05 (an ALFALFA object).
In the Wise Observatory images, no {\Halpha} emission was identified around the ALFALFA coordinates.
Within one arcminute from the ALFALFA coordinates of EIG 1s-05, all galaxies detected by SDSS have ${\SDSSg} > 21.6$, and none have spectroscopic redshifts. All GALEX detected objects in the same region have ${\magFUV} > 24$ and ${\magNUV} > 21$.
EIG 1s-05 may, therefore, be a ``dark galaxy'' with an extremely high HI to stellar mass ratio and a very low SFR. It may also be an ALFAFLA false detection, even though its SNR is 8.1 and it is considered a ``code 1'' object, i.e. a source of SNR and general qualities that make it a nearly {100\%} reliable detection \citep{2011AJ....142..170H}. 
}

\subsubsection*{EIG 1s-09}
\markChange
{
SDSS DR10 shows an edge-on galaxy, SDSS J112157.63+102959.6, {$\sim$13\,\arcsec} east of the centre of EIG 1s-09. The angular size of SDSS J112157.63+102959.6 is similar to that of EIG 1s-09. Its magnitude is ${\SDSSg} = 18.6$, compared to ${\SDSSg} = 16.9$ of EIG 1s-09. The redshift of SDSS J112157.63+102959.6 is unknown. Although there is a possibility that SDSS J112157.63+102959.6 is a close neighbour of EIG 1s-09, this seems unreasonable, since tidal tails are neither visible in the SDSS images nor in deeper images taken by the authors (Spector \& Brosch, in preparation).
}

\subsubsection*{EIG 1s-10}
\markChange
{
SDSS DR10 shows two objects at an angular distance of {$\sim$6\,\arcsec} from the centre of EIG 1s-10. One is north of EIG 1s-10, and is classified as a star by SDSS DR10. The second, classified as a galaxy, is south-west of EIG 1s-10. Both objects do not have measured redshifts. Although there is a possibility that one or both of these are galaxies merging with EIG 1s-10, this seems unreasonable, since tidal tails are neither visible in the SDSS images nor in deeper images taken by the authors (Spector \& Brosch, in preparation).
}

\subsubsection*{EIG 1s-11}
\markChange
{
The only redshift measurement found for EIG 1s-11 is from \cite{1993A&AS...98..275B} that quotes \cite{1987ApJS...63..247H}. This is a HI measurement made at the Arecibo observatory. The HI-profile for the galaxy was not published by \cite{1987ApJS...63..247H}. It is possible that the measurement ($4725 \pm 10$\,\kms) is a result of HI-confusion, and that EIG 1s-11 is actually a part of the Virgo cluster.
}

\subsubsection*{EIG 1a-02}
\markChange
{
SDSS DR10 shows a galaxy, SDSS J005629.17+241913.3, {$\sim$2\,\arcmin} west of EIG 1a-02 with unknown redshift. The angular size of SDSS J005629.17+241913.3 is not very different from that of EIG 1a-02. Its magnitude is ${\SDSSg} = 16.6$, compared to ${\SDSSg} = 17.0$ of EIG 1a-02.
Although there is a possibility that SDSS J005629.17+241913.3 is a close neighbour of EIG 1a-02, this seems unreasonable, since no tidal tails or other signs of interaction are visible in the SDSS images.
}

\subsubsection*{EIG 1a-04}
\markChange
{
{\Halpha} images of EIG 1a-04 showed strong star formation in LEDA 213033, a galaxy separated by {107\arcsec} from EIG 1a-04 (Spector \& Brosch, in preparation). Since LEDA 213033 has no measured redshift, its distance from EIG 1a-04 is unknown. The fact that it shows emission in the two narrow {\Halpha} filters used for the measurement (described in Spector \& Brosch, in preparation) indicates that its redshift is $\cz \cong 6000 \pm 1500\,\kms$. Therefore, the probability that it is less than {300\,\kms} away from EIG 1a-04 is estimated to be $\sim$10\%.
No sign of interaction between EIG 1a-04 and LEDA 213033 was detected.
}

\subsubsection*{EIG 3s-06}

\markChange
{
This is the only EIG that passes the isolation criterion using the ALFALFA dataset, but had neighbours closer than {3\,\Mpch} in the NED dataset. It was classified as part of subsample EIG-3s, because all of its NED neighbours are more than {2\,\Mpch} away from it.
}

\subsection{Galaxies found in the search but not included in the sample}
\label{App:EIGdata.Deleted}

\markChange
{
Two galaxies (CGCG 063-006 and UGC 09989) were first identified as EIGs but were later excluded from the sample, when neighbours closer than the {3\,\Mpch} limit were identified for them in SDSS.
In previous work \citep{2015PhDT} these galaxies were referred to as EIG 2s-03 and EIG 3s-08 (respectively).
For CGCG 063-006 a neighbour (SDSS J093248.11+121645.7) was found at a distance of {0.93\,\Mpc} in redshift space.
For UGC 09989 a neighbour (SDSS J154317.50+094155.8) was found at a distance of {1.24\,\Mpc} in redshift space. 
The visible images of UGC 09989 show an extension to the north-west, which ends in a galaxy of smaller angular size. This extension seems to be material extracted from the galaxy, possibly by interaction with a neighbour galaxy. WISE images show this extension clearly in bands W1 and W2 ({3.4\,\um} and {4.6\,\um} respectively).
}

Two objects (FGC 1647 and HIPASS J0835+14) were not included in the sample, because they were not found at their published coordinates. Data for FGC 1647 were published in the Flat Galaxy Catalogue \citep{1993AN....314...97K}. As advised by Igor Karachentsev, it does not appear in the Revised Flat Galaxy Catalogue (RFGC - \citealt{1999BSAO...47....5K}) possibly due to a mistake in coordinates. Data for HIPASS J0835+14 appears in the HI Parkes All Sky Survey (HIPASS) catalogue \cite{2006MNRAS.371.1855W}. This object was not included in the sample since no counterpart was found for it in the optical images or in the ALFALFA catalogue.

Three other objects (CGCG 043-113, SDSS J084236.58+103313.9 and SDSS J140626.67+092132.5) were not included in the sample after the redshift quoted for them in NED was found to be wrong, and using redshift obtained from alternative sources many close neighbours were found for them. For all three galaxies NED did not list sources for the redshift values it quotes. The accurate values obtained from alternative sources (SDSS DR10 for CGCG 043-113 and SDSS J084236.58+103313.9, and ALFALFA for SDSS J140626.67+092132.5) were significantly different.
Using these more accurate values CGCG 043-113 was found to have 28 neighbours in NED (and 23 in ALFALFA) closer than {3\,\Mpch}, with the closest neighbour at a distance of {0.38\,\Mpch}.
SDSS J084236.58+103313.9 was found to have three neighbours in NED (and five in ALFALFA) closer than {3\,\Mpch}.
SDSS J140626.67+092132.5 was found to have 65 neighbours in NED (and 24 in ALFALFA) closer than {3\,\Mpch}, with the closest neighbour at a distance of {0.15\,\Mpch}.

The following galaxies were not included in the sample because of their unreliable redshift values:
\begin{itemize}
  \item SDSS J104658.12+132911.3 -- For this galaxy NED quotes the value $3605 \pm 94\,\kms$ from SDSS DR5 \citep{2007ApJS..172..634A\rem{SDSS DR5}}, which is labelled with a warning status. Later SDSS releases list totally different values: $\z = 2.2 \pm 0.6$ with a warning status in SDSS DR7 \citep{2009ApJS..182..543A\rem{SDSS DR7}}, and $\z = 0.3 \pm 0.6$ in SDSS DR9 and DR10 \citep{2011AJ....142...72E\rem{SDSS DR9}, 2014ApJS..211...17A\rem{SDSS DR10}}. No other redshift measurement was found for this target in NED or in ALFALFA.

  \item {[}PGH98{]} 1228+1241 -- The uncertainty and source of the redshift measurement that NED quotes for this galaxy are not documented. The only reference that was found with a redshift measurement for this galaxy, \cite{1998Afz....41...32P}, quotes a significantly different value and does not list its uncertainty.

  \item VIII Zw 202 -- The only redshift measurement found for this galaxy is: $\cz = 3604 \pm 159\,\kms$ \citep{1996MNRAS.279..595D}. With this large uncertainty the isolation of the galaxy cannot be guaranteed.

  \item SDSS J134517.15+112452.6 -- SDSS is the only redshift source for this galaxy. SDSS DR7 measured $\cz = 5556 \pm 206\,\kms$, whereas SDSS DR8, DR9 and DR10 do not include its spectrum. With this uncertainty the isolation of the galaxy cannot be guaranteed.

  \item SDSS J153001.95+082550.9 -- For this galaxy NED quotes a redshift value from SDSS. However, different SDSS data releases measured significantly different values and issued warning statuses (SDSS DR6: $\cz = -141 \pm 110\,\kms$, SDSS DR7: $\cz = 2700 \pm 113\,\kms$ and SDSS DR10: $\cz = -79 \pm 15$). This could be due to a bright object close to the galaxy's line of sight that may contaminate the measured spectrum. No other redshift measurement was found for this target in NED or in ALFALFA.

  \item UCM 2241+2431 -- The only reference with a redshift measurement for this galaxy, \cite{1996AAS..120..323G}, does not quote its uncertainty. Therefore, the isolation of the galaxy cannot be guaranteed.

  \item 2MASX J23420930+2640174 -- For this galaxy NED quotes only one redshift measurement: $\cz = 6038 \pm 1001\,\kms$ \citep{1993ApJ...416...36P}. Neither ALFALFA nor SDSS measured a redshift for it. With this extreme uncertainty the isolation of the galaxy cannot be guaranteed, and its {\Halpha} emission measurement would not be accurate (because the transmittance of the {\Halpha} filter at the redshifted {\Halpha} line would not be known).

\end{itemize}